\documentclass[11pt,english]{article}
\usepackage{lmodern}
\usepackage[T1]{fontenc}
\usepackage[latin9]{inputenc}
\usepackage{xcolor}
\usepackage{amsmath}
\usepackage{amssymb}
\usepackage{cancel}
\usepackage{stackrel}
\usepackage{graphicx}
\usepackage{setspace}
\usepackage{wasysym}
\usepackage{colortbl}
\usepackage{multirow}
\usepackage{enumerate}
\makeatletter

\usepackage{jheppub}% for details on the use of the package, please
\usepackage{longtable}
\usepackage[font=small,labelfont=sc]{caption}
\usepackage[]{subcaption}
\usepackage{float}
\usepackage[compat=1.1.0]{tikz-feynman}
\tikzfeynmanset{warn luatex=false}
\usepackage[normalem]{ulem}
\usepackage{geometry}
\usepackage{booktabs}
\hypersetup{
linkcolor=red,
urlcolor=red,
citecolor=red,
linktocpage=true
}

\definecolor{darkred}{rgb}{0.6,0,0}
\definecolor{darkpurple}{rgb}{0.5,0,0.5}

\def\z2{$\mathbb{Z}_2$}
\def\z3{$\mathbb{Z}_3$}
\def\321{$SU(3)_c \times SU(2)_L \times U(1)_Y$}

\def\555{\ensuremath{SU(5)^3}}

\def\24{\ensuremath{\mathbf{24}}}

\def\SU15#1{\ensuremath{\mathbf{#1}}}

\definecolor{ForestGreen}{RGB}{34,139,34}

\providecommand{\tabularnewline}{\\}

\title{\boldmath Natural neutrino mass hierarchy in a theory of gauge flavour deconstruction}

\author[a]{Mario Fern\'andez Navarro,}
\author[b]{Stephen F. King}
\author[c,d]{and Avelino Vicente}
\affiliation[a]{School of Physics \& Astronomy, University of Glasgow, Glasgow G12 8QQ, UK}
\affiliation[b]{School of Physics \& Astronomy, University of Southampton, Southampton SO17 1BJ, UK}
\affiliation[c]{Instituto de F\'isica Corpuscular, CSIC-Universitat de Val\`encia, 46980 Paterna, Spain}
\affiliation[d]{Departament de F\'isica Te\`orica, Universitat de Val\`encia, 46100 Burjassot, Spain}

\emailAdd{Mario.FernandezNavarro@glasgow.ac.uk}
\emailAdd{S.F.King@soton.ac.uk}
\emailAdd{avelino.vicente@ific.uv.es}

\abstract{We show how a natural neutrino mass hierarchy with large lepton mixing angles may be achieved in a theory of gauge flavour deconstruction. Hitherto it has been shown that neutrino anarchy may result from such theories, but here we show that this need not necessarily be the case. In particular we consider the minimal tri-hypercharge theory, and show that the decomposition of the family hypercharges into the corresponding $B-L$ gauge groups,
together with the charged lepton mass hierarchy, implies the sequential dominance conditions for a neutrino mass hierarchy, where lepton mixing originates from both the neutrino and charged lepton sectors. We present novel and model-independent sequential dominance results applicable to this case, but also useful more generally. We also show how natural quark mass and mixing are included in such a framework.
}

\makeatother

\usepackage{babel}
\begin{document}
\makeatletter
\gdef\@fpheader{}
\makeatother

\maketitle \flushbottom

% Uncomment below for Introduction to start in new page
\clearpage
\pagenumbering{arabic}
\setcounter{page}{1}
%\numberwithin{equation}{section}
%\numberwithin{figure}{section}
%\counterwithout{equation}{section}
\allowdisplaybreaks

\section{Introduction}
\label{sec:intro}

The flavour problem remains one of the most intriguing puzzles of the Standard Model (SM), being
responsible for most of its parameters \cite{PDG:2024cfk}. Under the SM gauge group the three fermion families are identical, but differ greatly in mass showing a hierarchical pattern. The fact that quark mixing is small while lepton mixing is large only adds to the mystery.

An efficient mechanism to generate a hierarchical flavour structure consists of embedding the SM in a larger
gauge symmetry that contains a separate gauge group for each fermion family, with the light Higgs
doublet(s) originating from the third family group. This general idea was originally proposed in the early 80s \cite{Salam:1979p,Rajpoot:1980ib,Georgi:1981gj,Li:1981nk,Ma:1987ds,Ma:1988dn,Li:1992fi,Hill:1994hp,Barbieri:1994cx,Carone:1995ge,Muller:1996dj,Malkawi:1996fs,Dvali:2000ha,Asaka:2004ry,Babu:2007mb} and has received different names over the years, like ``tribal groups'', although the most recent literature is denoting this framework as \textit{flavour deconstruction} \cite{Craig:2011yk,Panico:2016ull,Bordone:2017bld,Greljo:2018tuh,Fuentes-Martin:2020bnh,Allwicher:2020esa,Fuentes-Martin:2020pww,Fuentes-Martin:2022xnb,Davighi:2022fer,Davighi:2022bqf,Davighi:2023iks,FernandezNavarro:2023rhv,Davighi:2023evx,FernandezNavarro:2023hrf,Davighi:2023xqn,Capdevila:2024gki,Barbieri:2023qpf,Fuentes-Martin:2024fpx,FernandezNavarro:2024hnv,Greljo:2024ovt,Covone:2024elw,Lizana:2024jby}. 

At the effective level, flavour-deconstructed models tend to generate small quark mixing, in agreement with the observed structure of the Cabibbo-Kobayashi-Maskawa (CKM) matrix. However, they tend to generate small lepton mixing as well. There are several ways to circumvent this issue and generate large Pontecorvo-Maki-Nakagawa-Sakata (PMNS) mixing, as required by oscillation data. One can go beyond the effective field theory (EFT), as in \cite{FernandezNavarro:2024hnv}, or one can also consider particular gauge symmetries where both hierarchical neutrino Yukawa couplings and hierarchical right-handed neutrino masses cancel the overall hierarchies in the effective neutrino mass matrix when applying the seesaw formula \cite{Greljo:2024ovt}. Alternatively, one can introduce extra scalar fields which only participate in the neutrino sector \cite{FernandezNavarro:2023rhv}, or charge all lepton doublets under the same family group, hence introducing gauge anomalies that need to be canceled by extra fermion content in the UV \cite{Fuentes-Martin:2024fpx,Lizana:2024jby}. However, all these mechanisms achieve an \textit{anarchical} pattern for the neutrino masses and mixing angles: the observed neutrino flavour structure is assumed to have originated as an accidental configuration of $\mathcal{O}(1)$ dimensionless coefficients. While this is a valid approach, it is widely speculated in the literature that the flavour structure of the lepton sector might not necessarily be of anarchic nature, but rather the simultaneous appearance of hierarchical
neutrino masses and two large mixing angles calls for further understanding, perhaps hinting at a particular dynamical mechanism.

Going in this direction, the framework of sequential dominance (SD) was proposed (see e.g.~\cite{King:1998jw,King:1999mb,King:2002nf,Antusch:2004gf}). Sequential dominance is not in itself a model, but a sub-mechanism within the general framework of the type-I seesaw, that may be applied to constructing different classes of neutrino models beyond anarchy. The starting point of sequential dominance is to assume that one of the right-handed neutrinos contributes dominantly to the heaviest neutrino mass, with the atmospheric
mixing angle being determined by a simple ratio of two Yukawa couplings. This is sometimes referred to as single right-handed neutrino dominance. Sequential dominance corresponds to the further assumption that, together with single right-handed neutrino dominance, a second right-handed neutrino contributes dominantly to the second heaviest neutrino mass, with the large solar mixing angle interpreted as a ratio of
Yukawa couplings. The third right-handed neutrino is effectively decoupled from the seesaw mechanism, and plays no part in determining the neutrino mass spectrum. If the decoupled right-handed neutrino is also the heaviest one, then sequential dominance is effectively equivalent to having two right-handed neutrinos.

The goal of this paper is to implement sequential dominance in the context of flavour deconstruction in order to go beyond anarchy in the neutrino sector. In particular, we will show that tri-hypercharge \cite{FernandezNavarro:2023rhv,FernandezNavarro:2024hnv}, which is arguably the simplest theory of flavour deconstruction, naturally delivers sequential dominance when extended to embrace the right-handed neutrino sector. We will see how the combination of flavour deconstruction with sequential dominance delivers non-trivial features, including the origin of PMNS mixing from both charged lepton and neutrino mixing: the atmospheric angle $\theta_{23}$ originates from both the 23 mixing angles of charged leptons and neutrinos, while the reactor angle $\theta_{13}$ originates mostly from a Cabibbo-like $\theta^{e}_{12}$ angle in the charged lepton sector, and the solar angle $\theta_{12}$ originates mostly from 12 neutrino mixing.

The  paper is structured as follows. In Section~\ref{sec:TH} we discuss the minimal tri-hypercharge theory as an example of anarchy in the neutrino sector of flavour deconstruction. In Section~\ref{sec:UVmodel} we show how sequential dominance naturally arises when tri-hypercharge is extended to embrace right-handed neutrinos, leading to potential predictivity in contrast to the hypothesis of anarchy. In this framework, we discuss the origin of a natural neutrino mass hierarchy, lepton mixing angles and a natural quark sector. Section~\ref{sec:conclusions} outlines our main conclusions. Our conventions are shown in Appendix~\ref{sec:conv}. Appendix~\ref{app:neutrinos} contains a review of sequential dominance in the neutrino sector, while Appendix \ref{app:charged_leptons} contains novel model-independent results regarding charged leptons in sequential dominance. Appendix~\ref{app:PMNS} contains model-independent formulas for the PMNS mixing angles that consider all potential contributions from neutrino and charged lepton mixing angles and phases. Finally, in Appendix~\ref{app:scalarPotential} we discuss the scalar potential for the considered model.

\section{Tri-hypercharge with anarchic neutrinos \label{sec:TH}}

The tri-hypercharge (TH) proposal is one of the simplest theories
of flavour deconstruction. It involves just assigning a separate
gauge hypercharge to each fermion family at high energies \cite{FernandezNavarro:2023rhv},
\begin{equation}
G_{\mathrm{TH}}=U(1)_{Y_{1}}\times U(1)_{Y_{2}}\times U(1)_{Y_{3}}\,,
\label{TH}
\end{equation}
which commutes with $SU(3)_{c}\times SU(2)_{L}$ that remains flavour universal as in the SM.
The TH symmetry is spontaneously broken down to the diagonal
hypercharge $U(1)_{Y}=U(1)_{Y_{1}+Y_{2}+Y_{3}}$
in two steps by a set of scalar fields charged under different hypercharges
which add to zero, denoted as ``hyperons''. This diagonal $U(1)_{Y}$ corresponds to the universal hypercharge of the SM. The Higgs doublet(s)
that spontaneously break electroweak symmetry are chosen to carry only
third family hypercharge, which allows to write only third family
Yukawa couplings at renormalisable level, while the Yukawa couplings of
the light families originate from non-renormalisable operators that
involve the hyperons. This successfully leads to a dynamical generation
of the observed charged fermion mass hierarchies and small quark mixing. In particular, we consider two
Higgs doublets $H_{u,d}$ that couple to up-quarks/neutrinos and down-quarks/charged leptons respectively\footnote{This may be enforced by particular mechanisms such as a softly broken $\mathbb{Z}_{2}$ symmetry, not specified here.}, in order to take into account the overall different
normalisation between the up sector and the down-quark/charged lepton
sectors.
Minimal ultraviolet (UV) completions of the non-renormalisable operators have been
provided via the inclusion of vector-like fermions and/or heavy Higgs
doublets \cite{FernandezNavarro:2024hnv}.

\begin{table}
\centering %
\begin{tabular}{lcccc}
\toprule 
Field  & $U(1)_{Y_{1}}$  & $U(1)_{Y_{2}}$  & $U(1)_{Y_{3}}$  & $SU(3)_{c}\times SU(2)_{L}$\tabularnewline
\midrule 
$\ell_{1}$  & $-\frac{1}{2}$  & 0  & 0  & ($\mathbf{1},\mathbf{2}$)\tabularnewline
$\ell_{2}$  & 0  & $-\frac{1}{2}$  & 0  & ($\mathbf{1},\mathbf{2}$)\tabularnewline
$\ell_{3}$  & 0  & 0  & $-\frac{1}{2}$  & ($\mathbf{1},\mathbf{2}$)\tabularnewline
\midrule 
$e_{1}^{c}$  & 1 & 0  & 0  & ($\mathbf{1},\mathbf{1}$)\tabularnewline
$e_{2}^{c}$  & 0  & 1  & 0  & ($\mathbf{1},\mathbf{1}$)\tabularnewline
$e_{3}^{c}$  & 0  & 0  & 1 & ($\mathbf{1},\mathbf{1}$)\tabularnewline
\midrule 
$\nu_{1}^{c}$  & 0  & 0  & 0  & ($\mathbf{1},\mathbf{1}$)\tabularnewline
$\nu_{2}^{c}$  & 0  & 0  & 0  & ($\mathbf{1},\mathbf{1}$)\tabularnewline
\midrule 
$H_{u}$  & 0  & 0  & $\frac{1}{2}$  & ($\mathbf{1},\mathbf{2}$)\tabularnewline
$H_{d}$  & 0  & 0  & $-\frac{1}{2}$  & ($\mathbf{1},\mathbf{2}$)\tabularnewline
\midrule 
$\phi_{12}$  & $\frac{1}{2}$  & $-\frac{1}{2}$  & 0  & ($\mathbf{1},\mathbf{1}$)\tabularnewline
$\phi_{23}$  & 0  & $\frac{1}{2}$  & $-\frac{1}{2}$  & ($\mathbf{1},\mathbf{1}$)\tabularnewline
\bottomrule
\end{tabular}\caption{Minimal tri-hypercharge model for the lepton sector \cite{FernandezNavarro:2024hnv}.
$H_{u,d}$ and $\phi_{ij}$ are scalars while the rest are the three
usual generations of chiral leptons plus two right-handed neutrinos.
\label{tab:Minimal-Tri-hypercharge-model}}
\end{table}

As shown in Table~\ref{tab:Minimal-Tri-hypercharge-model}, the minimal but complete realisation of the lepton sector \cite{FernandezNavarro:2024hnv}
involves only two hyperons, along with two right-handed neutrinos as full
singlets of the TH symmetry. Going beyond renormalisable level, this delivers the following Yukawa couplings
and Majorana masses for the neutrino singlets,
\begin{flalign}
\mathcal{L}=a_{3i}^{\nu} & \ell_{3}H_{u}\nu_{i}^{c}+a_{2i}^{\nu}\frac{\phi_{23}}{\Lambda^{\nu}_{23}}\ell_{2}H_{u}\nu_{i}^{c}+a_{1i}^{\nu}\frac{\phi_{12}}{\Lambda^{\nu}_{12}}\frac{\phi_{23}}{\Lambda^{\nu}_{23}}\ell_{1}H_{u}\nu_{i}^{c}+M_{ij}\nu_{i}^{c}\nu_{j}^{c}+\mathrm{h.c.}\,,\label{eq:tripe_EFT}
\end{flalign}
where $i=1,2$ and repeated indices are summed. Once the hyperons
get their VEVs, we obtain the following textures for the Dirac and
Majorana mass matrices of neutrinos, 
\begin{flalign}
\mathcal{L}_{\nu} & =\left(\begin{array}{ccc}
\ell_{1} & \ell_{2} & \ell_{3}\end{array}\right)m_{D}\left(\begin{array}{c}
\nu_{1}^{c}\\
\nu_{2}^{c}
\end{array}\right)+\left(\begin{array}{cc}
\nu_{1}^{c} & \nu_{2}^{c}\end{array}\right)M_{\mathrm{M}}\left(\begin{array}{c}
\nu_{1}^{c}\\
\nu_{2}^{c}
\end{array}\right)+\mathrm{h.c.}\\
 & =\left(\begin{array}{ccc}
\ell_{1} & \ell_{2} & \ell_{3}\end{array}\right)\left(\begin{array}{cc}
a_{11}^{\nu}\epsilon_{12}^{\nu}\epsilon_{23}^{\nu} & a_{12}^{\nu}\epsilon_{12}^{\nu}\epsilon_{23}^{\nu}\\
a_{21}^{\nu}\epsilon_{23}^{\nu} & a_{22}^{\nu}\epsilon_{23}^{\nu}\\
a_{31}^{\nu} & a_{32}^{\nu}
\end{array}\right)\left(\begin{array}{c}
\nu_{1}^{c}\\
\nu_{2}^{c}
\end{array}\right)H_{u}+\left(\begin{array}{cc}
\nu_{1}^{c} & \nu_{2}^{c}\end{array}\right)\left(\begin{array}{cc}
M_{22} & M_{23}\\
M_{32} & M_{33}
\end{array}\right)\left(\begin{array}{c}
\nu_{1}^{c}\\
\nu_{2}^{c}
\end{array}\right)+\mathrm{h.c.}\,,\label{eq:MassMatrices_tripe}
\end{flalign}
where we have defined $\epsilon_{12}^{\nu}=\langle\phi_{12}\rangle/\Lambda_{12}^{\nu}$
and $\epsilon_{23}^{\nu}=\langle\phi_{23}\rangle/\Lambda_{23}^{\nu}$.
The effective mass matrix for active neutrinos is obtained after applying
the seesaw formula,
\begin{equation}
m_{\nu}\simeq m_{D}(M_{\mathrm{M}})^{-1}m_{D}^{\mathrm{T}}\,.
\end{equation}
In the
charged lepton sector, we obtain the following Yukawa couplings,
\begin{equation}
\mathcal{L}_{e}=\left(\begin{array}{ccc}
\ell_{1} & \ell_{2} & \ell_{3}\end{array}\right)\left(\begin{array}{ccc}
a_{11}^{e}\epsilon_{12}^{e}\epsilon_{23}^{e} & a_{12}^{e}\epsilon_{12}^{e}\epsilon_{23}^{e} & a_{13}^{e}\epsilon_{12}^{e}\epsilon_{23}^{e}\\
a_{21}^{e}(\epsilon_{12}^{e}){}^{2}\epsilon_{23}^{e} & a_{22}^{e}\epsilon_{23}^{e} & a_{23}^{e}\epsilon_{23}^{e}\\
a_{31}^{e}(\epsilon_{12}^{e}){}^{2}(\epsilon_{23}^{e}){}^{2} & a_{32}^{e}(\epsilon_{23}^{e}){}^{2} & a_{33}^{e}
\end{array}\right)\left(\begin{array}{c}
e_{1}^{c}\\
e_{2}^{c}\\
e_{3}^{c}
\end{array}\right)H_{d}+\mathrm{h.c.}\,,\label{eq:Yukawa_e_tripe}
\end{equation}
where we have defined $\epsilon^{e}_{12}=\langle\phi_{12}\rangle/\Lambda^{e}_{12}$
and $\epsilon^{e}_{23}=\langle\phi_{23}\rangle/\Lambda^{e}_{23}$.
Under the
validity of the EFT introduced in Eq.~(\ref{eq:tripe_EFT}), we expect
$\epsilon^{\nu,e}_{12,23}\ll1$ to imprint hierarchies among the rows of the Yukawa couplings,
while $M_{ij}$ are naturally presumed to be of similar order since no symmetry distinguishes the two singlet neutrinos. This
is a general result in theories of flavour deconstruction (see e.g.~the discussion of \cite{Greljo:2024ovt}), where one
naively expects a hierarchical lepton sector with small mixing angles,
similar to the observed pattern of the quark sector, unless the dimensionless coefficients
$a^{\nu,e}$ are fine-tuned. This is however in contradiction with the
two large mixing angles observed in the PMNS matrix.

So far, the four possibilities to generate large lepton mixing in theories
of flavour deconstruction consist on:
\begin{enumerate}[(i)]
\item Introduce extra linking scalars (e.g.~hyperons) which only participate in the neutrino sector and change the Yukawa texture above \cite{FernandezNavarro:2023rhv}.
\item Go beyond the validity of the EFT approach
of Eq.~(\ref{eq:tripe_EFT}) to generate $\epsilon^{\nu}_{12,23}\sim1$
in the full UV theory \cite{FernandezNavarro:2024hnv}.
\item Consider particular gauge symmetries
where both hierarchical neutrino Yukawa couplings and hierarchical
right-handed neutrino masses cancel the overall hierarchies in the
neutrino mass matrix when applying the seesaw formula \cite{Greljo:2024ovt}.
The extended gauge theory will then distinguish between the right-handed neutrinos.
\item Charge all lepton doublets under the same site (or hypercharge). This introduces gauge anomalies that can be canceled with extra fermionic content in the UV \cite{Fuentes-Martin:2024fpx,Lizana:2024jby}.
\end{enumerate}
In all these cases, one obtains an \textit{anarchic} effective neutrino
mass matrix: all entries are governed by $\mathcal{O}(1)$ coefficients
which are simply fitted to neutrino oscillation data. This is successful at the
level of reconciling the hypothesis of flavour deconstruction with
neutrino data, but it does not give any understanding about the observed
neutrino flavour structure, which is assumed to have originated as
an accidental configuration of $\mathcal{O}(1)$ dimensionless coefficients.
However, the flavour
structure of the lepton sector might not necessarily be anarchic
in nature. This would be the case if the physical neutrino masses exhibited a strong hierarchy, where one of the right-handed neutrinos dominantly contributes to the heaviest physical neutrino mass, as in sequential
dominance~\cite{King:1998jw,King:1999mb,King:2002nf,Antusch:2004gf}.
The simultaneous appearance of hierarchical neutrino
masses and two large (and one small) mixing angles may thus hint towards a dynamical mechanism beyond anarchy. 

Such a dynamical mechanism requires us to go beyond tri-hypercharge.
The reason why this is necessary is that the two right-handed neutrinos are both singlets under the tri-hypercharge gauge group, and hence are indistinguishable, which results in the two columns of the Dirac matrix being approximately equal, and the heavy Majorana mass matrix being anarchical. 
In the current framework it is therefore difficult to obtain a natural neutrino mass hierarchy, and this motivates extending tri-hypercharge to a larger gauge group, under which the right-handed neutrinos are no longer indistinguishable singlets.

In the following section, we will show that the conditions of sequential dominance 
(see Appendix~\ref{sec:seq}), which naturally provides a first step to
understand the dynamic origin of the observed pattern of neutrino
data, are naturally realised when the tri-hypercharge setup
is minimally extended such that the right-handed neutrinos are no longer indistinguishable singlets. In this way,
it is possible to go beyond anarchy in the neutrino sector of flavour deconstruction.

\section{Natural neutrino mass hierarchy from UV extension of tri-hypercharge} \label{sec:UVmodel}
It is well known that a natural neutrino mass hierarchy can result from the sequential dominance of three right-handed neutrinos (see e.g.~\cite{King:1998jw,King:1999mb,King:2002nf,Antusch:2004gf}), where one of them is decoupled and may be ignored, while the other two contribute sequentially to the seesaw mechanism. In TH theories, this may be achieved by extending two of the hypercharge gauge groups into respective $B-L$ gauge groups, which prevents large Majorana masses, while one hypercharge gauge group remains intact, allowing one heavy decoupled right-handed neutrino. That is the strategy that we follow in this section.

\subsection{Gauge symmetry and symmetry breaking}

\begin{table}
\centering{}%
\begin{tabular}{lcccc}
\toprule 
Field  & $U(1)_{Y_{1}}$  & $U(1)_{R_{2}}\times U(1)_{(B-L)_{2}/2}$  & $U(1)_{R_{3}}\times U(1)_{(B-L)_{3}/2}$ & $SU(3)_{c}\times SU(2)_{L}$\tabularnewline
\midrule 
$\ell_{1}$  & $-\frac{1}{2}$  & $\mathrm{(0,0)}$  & $\mathrm{(0,0)}$ & ($\mathbf{1},\mathbf{2}$)\tabularnewline
$\ell_{2}$  & 0  & $\mathrm{(0,-\frac{1}{2})}$  & $\mathrm{(0,0)}$ & ($\mathbf{1},\mathbf{2}$)\tabularnewline
$\ell_{3}$  & 0  & $\mathrm{(0,0)}$  & $\mathrm{(0,-\frac{1}{2})}$ & ($\mathbf{1},\mathbf{2}$)\tabularnewline
\midrule 
$e_{1}^{c}$  & 1  & $\mathrm{(0,0)}$  & $\mathrm{(0,0)}$ & ($\mathbf{1},\mathbf{1}$)\tabularnewline
$e_{2}^{c}$  & 0  & $\mathrm{(\frac{1}{2},\frac{1}{2})}$  & $\mathrm{(0,0)}$ & ($\mathbf{1},\mathbf{1}$)\tabularnewline
$e_{3}^{c}$  & 0  & $\mathrm{(0,0)}$  & $\mathrm{(\frac{1}{2},\frac{1}{2})}$ & ($\mathbf{1},\mathbf{1}$)\tabularnewline
\midrule 
$\nu_{2}^{c}$  & 0  & $\mathrm{(-\frac{1}{2},\frac{1}{2})}$  & $\mathrm{(0,0)}$ & ($\mathbf{1},\mathbf{1}$)\tabularnewline
$\nu_{3}^{c}$  & 0  & $\mathrm{(0,0)}$  & $\mathrm{(-\frac{1}{2},\frac{1}{2})}$ & ($\mathbf{1},\mathbf{1}$)\tabularnewline
\midrule 
$H_{u,d}$  & 0  & $\mathrm{(0,0)}$  & $\mathrm{(\pm\frac{1}{2},0)}$ & ($\mathbf{1},\mathbf{2}$)\tabularnewline
\midrule 
$\chi_{2}$  & 0  & $\mathrm{(1,-1)}$  & $\mathrm{(0,0)}$ & ($\mathbf{1},\mathbf{1}$)\tabularnewline
$\chi_{3}$  & 0  & $\mathrm{(0,0)}$  & $\mathrm{(1,-1)}$ & ($\mathbf{1},\mathbf{1}$)\tabularnewline
\midrule 
$\phi_{12}^{R}$  & $\frac{1}{2}$  & $\mathrm{(-\frac{1}{2},0)}$  & $\mathrm{(0,0)}$ & ($\mathbf{1},\mathbf{1}$)\tabularnewline
$\phi_{12}^{L}$  & $\frac{1}{2}$  & $\mathrm{(0,-\frac{1}{2})}$  & $\mathrm{(0,0)}$ & ($\mathbf{1},\mathbf{1}$)\tabularnewline
$\phi_{23}^{R}$  & 0  & $\mathrm{(\frac{1}{2},0)}$  & $\mathrm{(-\frac{1}{2},0)}$ & ($\mathbf{1},\mathbf{1}$)\tabularnewline
$\phi_{23}^{L}$  & 0  & $\mathrm{(0,\frac{1}{2})}$  & $\mathrm{(0,-\frac{1}{2})}$ & ($\mathbf{1},\mathbf{1}$)\tabularnewline
\bottomrule
\end{tabular}\caption{Field content relevant for the lepton sector. $H_{u,d}$, $\chi_{2,3}$
and $\phi_{ij}^{R,L}$ are scalars while the rest are the three usual
generations of chiral leptons plus two right-handed neutrinos. \label{tab:Lepton}}
\end{table}

In this section, we will show that the conditions of sequential
dominance are naturally achieved when tri-hypercharge
is extended to a larger gauge group under which two of the $\nu_{i}^{c}$ neutrinos
are not singlets, thereby preventing their Majorana masses until the larger group is broken. To this end, we consider the tri-hypercharge
gauge symmetry in Eq.~\eqref{TH} to be an effective low energy theory resulting from the ultraviolet (UV) gauge group, 
\begin{equation}
G_{\mathrm{UV}}=U(1)_{Y_{1}}\times U(1)_{R_{2}}\times U(1)_{(B-L)_{2}/2}\times U(1)_{R_{3}}\times U(1)_{(B-L)_{3}/2}\,,
\label{UV}
\end{equation}
while $SU(3)_{c}$ and $SU(2)_{L}$ remain universal. The decomposition
of hypercharge into an Abelian symmetry for right-handed particles
$U(1)_{R}$ and baryon-minus-lepton number $U(1)_{B-L}$, is well
motivated from left-right symmetric embeddings of the SM, such as
the Pati-Salam \cite{Pati:1974yy} and $SO(10)$ \cite{Fritzsch:1974nn,Georgi:1974my} theories. Here we apply this idea
to the second and third family hypercharges, 
with the gauge group in Eq.~\eqref{UV} broken down to the TH gauge group in Eq.~\eqref{TH} at high energy, as discussed below, where the hypercharge generators are given by
\begin{align}
    Y_{2}&=R_{2}+\frac{1}{2}(B-L)_{2} \, , \\
    Y_{3}&=R_{3}+\frac{1}{2}(B-L)_{3} \, ,
\end{align}
noting
the $1/2$ normalisation in the $B-L$ charge definition to be consistent
with our normalisation of hypercharge.

Two right-handed neutrinos $\nu_{2,3}^{c}$ are now required by anomaly
cancellation, and they are charged under $U(1)_{R_{2}}\times U(1)_{(B-L)_{2}/2}$
and $U(1)_{R_{3}}\times U(1)_{(B-L)_{3}/2}$ respectively. In general,
we may also decompose $U(1)_{Y_{1}}$ into $R_{1}$ and $(B-L)_{1}$,
providing another right-handed neutrino $\nu_{1}^{c}$ from anomaly
cancellation. However, as we will show in the following, our model
will naturally generate sequential dominance, where $\nu_{1}^{c}$
provides subleading\footnote{In particular, any contribution from $\nu_{1}^{c}$ will be higher
order in the sequential dominance expansion.
Note that we are assuming only mild hierarchies of dimensionless couplings and Majorana masses, which necessitates the gauge extension discussed here, since by inspecting the mass matrix textures in Eq.~(\ref{eq:MassMatrices_tripe})
and the sequential dominance conditions in Appendix~\ref{sec:seq}, it is clear that unextended
tri-hypercharge does not generate sequential dominance unless the
dimensionless coefficients or the Majorana mass terms are assumed
to be hierarchical. 
} contributions to the seesaw mechanism. Therefore, for simplicity
we will assume that $Y_{1}$ is not decomposed until higher scales
so that $\nu_{1}^{c}$ is effectively decoupled from the seesaw.

In order to break the symmetry, we introduce the scalars $\chi_{i}\sim(1,-1)_{i}$, with $i=2,3$
as shown in Table~\ref{tab:Lepton}, whose VEVs spontaneously
break the UV group in Eq.~\eqref{UV} down to the TH group in Eq.~\eqref{TH},
\begin{equation}
G_{\mathrm{UV}}\overset{\langle\chi_{2,3}\rangle}{\longrightarrow}G_{\mathrm{TH}}\,,
\end{equation}
corresponding to 
$U(1)_{R_{i}}\times U(1)_{(B-L)_{i}/2}\rightarrow U(1)_{Y_{i}}$ thereby 
recovering tri-hypercharge,
simultaneously allowing Majorana masses for the right-handed neutrinos $m_{\nu_{i}^{c}}\sim\langle\chi_{i}\rangle$.
We expect these VEVs to be large enough to provide
most of the suppression for small neutrino masses, in a natural framework. 

Finally, we also
introduce the scalars $\phi_{12,23}^{R,L}$ as shown in Table~\ref{tab:Lepton},
which play the role of the hyperons in the TH theory. Indeed after
the $\chi_{i}$ scalars get their VEVs, the scalars $\phi_{12,23}^{R,L}$
reduce to the hyperons $\phi_{12,23}$ of TH as shown in Table~\ref{tab:Minimal-Tri-hypercharge-model}.
They get VEVs \cite{FernandezNavarro:2023rhv,FernandezNavarro:2024hnv} which break tri-hypercharge down to the SM, i.e.~the chain of symmetry breaking is
\begin{equation}
G_{\mathrm{UV}}\overset{\langle\chi_{i}\rangle}{\rightarrow}G_{\mathrm{TH}}\overset{\langle\phi_{ij}\rangle}{\rightarrow}U(1)_{Y}\,,
\end{equation}
where $\langle\chi_{i}\rangle > \langle\phi_{ij}\rangle$
and $\langle\chi_{i}\rangle$ is large enough to provide the required suppression for neutrino masses, as we shall see. We also need to consider Higgs
doublet(s) to break spontaneously the electroweak symmetry. Following the same strategy as in 
the original implementation of tri-hypercharge~\cite{FernandezNavarro:2023rhv}, we consider two Higgs doublets
$H_{u,d}$ to take into account the overall different normalisation
between the up sector and the down-quark/charged lepton sectors. These
are only charged under $R_{3}$ to allow for third family renormalisable
Yukawa couplings only, as in tri-hypercharge. The scalar potential is discussed in Appendix \ref{app:scalarPotential}, where we also show that all Goldstone modes may be given heavy masses.

\subsection{Leptons}

We start by writing the Yukawa couplings involving charged leptons,
along with the Dirac and Majorana mass matrices in the neutrino sector\footnote{Without loss of generality, we absorb the Yukawa couplings $a^{\nu^{c}}_{22}$ and $a^{\nu^{c}}_{33}$ into the definitions of $\langle \chi_{2}\rangle$ and $\langle \chi_{3}\rangle$ in $M_{\mathrm{M}}$.},
\begin{equation}
\mathcal{L}_{e}=\left(\begin{array}{ccc}
\ell_{1} & \ell_{2} & \ell_{3}\end{array}\right)\left(\begin{array}{ccc}
a_{11}^{e}\epsilon_{12}^{R}\epsilon_{23}^{R} & a_{12}^{e}\epsilon_{12}^{L}\epsilon_{23}^{R} & a_{13}^{e}\epsilon_{12}^{L}\epsilon_{23}^{L}\\
a_{21}^{e}\epsilon_{12}^{L}\epsilon_{12}^{R}\epsilon_{23}^{R} & a_{22}^{e}\epsilon_{23}^{R} & a_{23}^{e}\epsilon_{23}^{L}\\
a_{31}^{e}\epsilon_{12}^{L}\epsilon_{12}^{R}\epsilon_{23}^{L}\epsilon_{23}^{R} & a_{32}^{e}\epsilon_{23}^{L}\epsilon_{23}^{R} & a_{33}^{e}
\end{array}\right)\left(\begin{array}{c}
e_{1}^{c}\\
e_{2}^{c}\\
e_{3}^{c}
\end{array}\right)H_{d}+\mathrm{h.c.}\,,\label{eq:Yukawa_e}
\end{equation}
\begin{flalign}
\mathcal{L}_{\nu} & =\left(\begin{array}{ccc}
\ell_{1} & \ell_{2} & \ell_{3}\end{array}\right)m_{D}\left(\begin{array}{c}
\nu_{1}^{c}\\
\nu_{2}^{c}
\end{array}\right)+\left(\begin{array}{cc}
\nu_{1}^{c} & \nu_{2}^{c}\end{array}\right)M_{\mathrm{M}}\left(\begin{array}{c}
\nu_{1}^{c}\\
\nu_{2}^{c}
\end{array}\right)+\mathrm{h.c.}\label{eq:Neutrino_textures}\\
 & =\left(\begin{array}{ccc}
\ell_{1} & \ell_{2} & \ell_{3}\end{array}\right)\left(\begin{array}{cc}
a_{12}^{\nu}\epsilon_{12}^{L}\epsilon_{23}^{R} & a_{13}^{\nu}\epsilon_{12}^{L}\epsilon_{23}^{L}\\
a_{22}^{\nu}\epsilon_{23}^{R} & a_{23}^{\nu}\epsilon_{23}^{L}\\
a_{32}^{\nu}\epsilon_{23}^{L}\epsilon_{23}^{R} & a_{33}^{\nu}
\end{array}\right)\left(\begin{array}{c}
\nu_{1}^{c}\\
\nu_{2}^{c}
\end{array}\right)H_{u}+\left(\begin{array}{cc}
\nu_{1}^{c} & \nu_{2}^{c}\end{array}\right)\left(\begin{array}{cc}
\chi_{2} & \epsilon_{23}^{L}\epsilon_{23}^{R}\chi_{3}\\
\epsilon_{23}^{L}\epsilon_{23}^{R}\chi_{3} & \chi_{3}
\end{array}\right)\left(\begin{array}{c}
\nu_{1}^{c}\\
\nu_{2}^{c}
\end{array}\right)+\mathrm{h.c.}\nonumber 
\end{flalign}
where $\epsilon_{ij}^{R,L}=\langle\phi_{ij}^{R,L}\rangle/\Lambda_{ij}$
and $a_{ij}^{e,\nu}$ are dimensionless coefficients expected to be
of $\mathcal{O}(1)$.

The charged lepton texture above delivers the following approximate
scalings for the charged leptons mass eigenvalues, 
\begin{equation}
m_{e}\sim\epsilon_{12}^{R}\epsilon_{23}^{R}\langle H_{d}\rangle\,,\quad m_{\mu}\sim\epsilon_{23}^{R}\langle H_{d}\rangle\,,\quad m_{\tau}\sim\langle H_{d}\rangle.
\end{equation}
Therefore, we obtain approximate numerical values for the small parameters
$\epsilon_{12,23}^{R}$ in order to reproduce the observed charged
lepton mass spectrum, 
\begin{equation}
\epsilon_{12}^{R}\sim\frac{m_{e}}{m_{\mu}}\simeq0.005\,,\qquad\epsilon_{23}^{R}\sim\frac{m_{\mu}}{m_{\tau}}\simeq0.06\,.
\end{equation}

The coefficients $\epsilon_{12,23}^{L}$ are relevant for charged
lepton and neutrino mixing, which ultimately will contribute to the
PMNS mixing angles. Therefore, we expect $\epsilon_{ij}^{R}<\epsilon_{ij}^{L}$
motivated from the observed patterns of charged lepton masses and
PMNS mixing. This imprints a hierarchical column structure into the
Yukawa couplings of charged leptons and neutrinos, which suggests
the presence of sequential dominance \cite{King:1998jw,King:1999mb,King:2002nf,Antusch:2004gf}. In particular, in our
model the conditions of sequential dominance in the neutrino and charged
lepton sectors (see Eqs.~(\ref{eq:SD_nu}) and (\ref{eq:SD_charged_lepton}))
translate into 
\begin{equation}
\frac{|a_{33}^{\nu}|^{2},|a_{23}^{\nu}\epsilon_{23}^{L}|^{2},|a_{33}^{\nu}a_{23}^{\nu}\epsilon_{23}^{L}|}{\langle\chi_{3}\rangle}\gg\frac{|x^{\nu}y^{\nu}|}{\langle\chi_{2}\rangle}\,,\label{eq:SD_neutrinos_model}
\end{equation}
\begin{equation}
|a_{33}^{e}|^{2},|a_{23}^{e}\epsilon_{23}^{L}|^{2},|a_{33}^{e}a_{23}^{e}\epsilon_{23}^{L}|\gg|x^{e}y^{e}|\gg|(x^{e})'(y^{e})'|\,,\label{eq:SD_chargedlepton_model}
\end{equation}
where 
\begin{equation}
x^{\nu,e},\,y^{\nu,e}=a_{32}^{\nu,e}\epsilon_{23}^{R}\epsilon_{23}^{L},\,a_{22}^{\nu,e}\epsilon_{23}^{R},\,a_{12}^{\nu.e}\epsilon_{23}^{R}\epsilon_{12}^{L}\,,
\end{equation}
\begin{equation}
(x^{e})',\,(y^{e})'=a_{11}^{e}\epsilon_{12}^{R}\epsilon_{23}^{R},\,a_{21}^{e}\epsilon_{12}^{L}\epsilon_{12}^{R}\epsilon_{23}^{R},\,a_{31}^{e}\epsilon_{12}^{L}\epsilon_{23}^{L}\epsilon_{12}^{R}\epsilon_{23}^{R}\,.
\end{equation}
The presence of sequential dominance allows us to derive simple analytical
formulas for mass eigenvalues and mixing angles at leading order in
the sequential dominance expansion, following the results in Appendix~\ref{sec:seq}.
For the neutrino sector, we obtain 
\begin{flalign}
\tan\theta_{23}^{\nu} & \simeq\frac{|a_{23}^{\nu}|}{|a_{33}^{\nu}|}\epsilon_{23}^{L}\,,\\
\theta_{13}^{\nu} & \simeq\frac{|a_{13}^{\nu}|}{\sqrt{|a_{33}^{\nu}|^{2}+|a_{23}^{\nu}\epsilon_{23}^{L}|^{2}}}\epsilon_{12}^{L}\epsilon_{23}^{L}\,,\\
\tan\theta_{12}^{\nu} & \simeq\frac{\left|a_{12}^{\nu}\right|\epsilon_{12}^{L}}{c_{23}^{\nu}\left|a_{22}^{\nu}\right|\cos(\tilde{\phi}_{a_{22}^{\nu}})-s_{23}^{\nu}|a_{32}^{\nu}|\cos(\tilde{\phi}_{a_{32}^{\nu}})\epsilon_{23}^{L}}\,,\\
m_{3} & \simeq\frac{|a_{33}^{\nu}|^{2}+|a_{23}^{\nu}\epsilon_{23}^{L}|^{2}}{\langle\chi_{3}\rangle}\langle H_{u}\rangle^{2}\,,\label{eq:m3_model}\\
m_{2} & \simeq\frac{\left|a_{12}^{\nu}\epsilon_{12}^{L}\epsilon_{23}^{R}\right|^{2}}{\langle\chi_{2}\rangle(s_{12}^{\nu})^{2}}\langle H_{u}\rangle^{2}\,,\label{eq:m2_model}\\
m_{1} & =0\,,
\end{flalign}
while for the charged lepton sector we obtain 
\begin{flalign}
\tan\theta_{23}^{e} & \simeq\frac{|a_{23}^{e}|}{|a_{33}^{e}|}\epsilon_{23}^{L}\,,\\
\theta_{13}^{e} & \simeq\frac{|a_{13}^{e}|}{\sqrt{|a_{33}^{e}|^{2}+|a_{23}^{e}\epsilon_{23}^{L}|^{2}}}\epsilon_{12}^{L}\epsilon_{23}^{L}\,,\\
\tan\theta_{12}^{e} & \simeq\frac{\left|a_{12}^{e}\right|\epsilon_{12}^{L}}{c_{23}^{e}\left|a_{22}^{e}\right|\cos(\tilde{\phi}_{a_{12}^{e}})-s_{23}^{e}|a_{32}^{e}|\cos(\tilde{\phi}_{a_{32}^{e}})\epsilon_{23}^{L}}\,,\\
m_{\tau} & \simeq\sqrt{|a_{33}^{e}|^{2}+|a_{23}^{e}\epsilon_{23}^{L}|^{2}}\langle H_{d}\rangle\,,\\
m_{\mu} & \simeq\frac{\left|a_{12}^{e}\epsilon_{12}^{L}\epsilon_{23}^{R}\right|}{s_{12}^{e}}\langle H_{d}\rangle\,,\\
m_{e} & \simeq\left[|a_{11}^{e}|c_{12}^{e}\cos(\tilde{\phi}_{a_{11}^{e}})-|a_{21}^{e}\epsilon_{12}^{L}|s_{12}^{e}c_{23}^{e}\cos(\tilde{\phi}_{a_{21}^{e}})\right.\\
 & \left.+|a_{31}^{e}\epsilon_{12}^{L}\epsilon_{23}^{L}|s_{12}^{e}s_{23}^{e}\cos(\tilde{\phi}_{a_{31}^{e}})\right]\epsilon_{12}^{R}\epsilon_{23}^{R}\langle H_{d}\rangle\,,
\end{flalign}
where the phases $\tilde{\phi}$ are defined in Appendix~\ref{sec:seq}.

The sequential dominance conditions in the charged lepton sector (\ref{eq:SD_chargedlepton_model})
are automatically satisfied by the hierarchical column structure of
the Yukawa textures, enforced by the smaller $\epsilon_{12,23}^{R}$
and larger $\epsilon_{12,23}^{L}$, which arise as a natural consequence
of explaining the charged lepton mass hierarchies and large lepton
mixing simultaneously. In the neutrino sector, the strong hierarchy
by columns also suggests sequential dominance, although the conditions
(\ref{eq:SD_neutrinos_model}) also depend on $\langle\chi_{2}\rangle$
and $\langle\chi_{3}\rangle$. Due to the $\epsilon_{23}^{L}\epsilon_{23}^{R}$
suppression of the off-diagonal entries in $M_{\mathrm{M}}$, we shall
approximate this as a diagonal matrix. By using the formulas (\ref{eq:m3_model})
and (\ref{eq:m2_model}) with $\epsilon_{23}^{R}\sim0.06$, $\epsilon_{12,23}^{L}\gtrsim0.1$
and setting the dimensionless coefficients to $\mathcal{O}(1)$, we
find that in order to reproduce neutrino data we have $\langle\chi_{2}\rangle\sim10^{13}\,\mathrm{GeV}$
and $\langle\chi_{3}\rangle\sim10^{14}\,\mathrm{GeV}$. Departing
from these values would require tuning of dimensionless coefficients
in order to be consistent with lepton data. For consistency of the
EFT we assume the cutoffs $\Lambda_{ij}$ to be the largest scales
of the theory, i.e.~$\Lambda_{ij}\gtrsim\langle\chi_{i}\rangle\,,\langle\phi_{ij}^{R,L}\rangle$.
This implies that the VEVs $\langle\phi_{ij}^{R,L}\rangle$ are high
scale\footnote{We note that by going beyond the EFT and specifying the degrees of
freedom corresponding to $\Lambda_{ij}$ as in \cite{FernandezNavarro:2024hnv},
a hierarchy of scales $\langle\chi_{i}\rangle\gg\langle\phi_{ij}^{R,L}\rangle$
would be possible without changing our conclusions, and the tri-hypercharge
symmetry breaking may be realised at the TeV scale $\langle\phi_{ij}^{R,L}\rangle\sim\mathcal{O}(\mathrm{TeV})$,
with the associated low energy phenomenology~\cite{FernandezNavarro:2023rhv,FernandezNavarro:2024hnv}. Assuming a UV completion involving vector-like (VL) fermions, this phenomenology includes charged lepton flavour violating (CLFV) processes mediated by $Z'$ gauge bosons at tree-level. At 1-loop there would be diagrams involving $Z'$ and VL charged leptons (see Figure 5 in \cite{FernandezNavarro:2024hnv}) along with diagrams involving $W$ bosons and VL neutrinos. These processes may reveal information on the sequential dominance structure in both charged leptons and neutrino sectors, along the lines of~\cite{Blazek:2002wq}.}, most likely $\langle\phi_{23}^{R,L}\rangle\gtrsim10^{13}\,\mathrm{GeV}$
and $\langle\phi_{12}^{R,L}\rangle\gtrsim10^{12}\,\mathrm{GeV}$.
The very heavy scales $\langle\chi_{2}\rangle$ and $\langle\chi_{3}\rangle$
are then responsible for the smallness of active neutrino masses in
our model. The fact that both are close to each other may suggest
the presence of a cyclic symmetry relating the three families and
providing gauge unification of our model in a flavour deconstructed
$SO(10)^{3}$ framework, similar to tri-unification in $SU(5)^{3}$
\cite{FernandezNavarro:2023hrf}. 

Moreover, the presence of sequential dominance in our model translates
into potential predictivity over the origin of the neutrino flavour pattern:
\begin{enumerate}
\item We obtain a natural mass hierarchy among normally ordered neutrino
mass eigenvalues, 
\begin{equation}
m_{3}^{2}\gg m_{2}^{2}\gg m_{1}^{2}\,.
\end{equation}
\item The atmospheric neutrino mass $m_{3}$ and the mixing angle $\theta_{23}^{\nu}$
are determined by the couplings of the dominant right-handed neutrino
with mass $\langle\chi_{3}\rangle$ . The solar neutrino mass $m_{2}$
and the mixing angle $\theta_{12}^{\nu}$ are determined by the couplings
of the subdominant right-handed neutrino of mass $\langle\chi_{2}\rangle$.
Moreover, these mixing angles and mass eigenvalues are described by the 
simple analytical formulas given above, at leading order in the sequential dominance
expansion, which were derived from the general results in  Appendix~\ref{app:neutrinos}.
\item In our model the lightest neutrino is massless, $m_{1} = 0$,
in good agreement with the current bounds on the neutrino scale $\Sigma m_{\nu}$
by cosmological observations~\cite{DESI:2024mwx} and by the KATRIN experiment~\cite{KATRIN:2024cdt}. However, a tiny mass will be generated if we decompose $Y_{1}$
as discussed before, connected to a heavier and effectively decoupled
right-handed neutrino with mass $\langle\chi_{1}\rangle$. This would
be necessary for the potential embedding into a grand unified scenario, such as $SO(10)^{3}$.
\item Sequential dominance in the charged lepton sector as delivered by our Yukawa texture ensures a naturally hierarchical mass spectrum and suppressed right-handed charged lepton mixing. Mass eigenvalues and mixing angles are described by the simple compact formulas above,
derived from the model-independent results in Appendix~\ref{app:charged_leptons} at leading order in the sequential dominance expansion.
\end{enumerate}
This potential predictivity can be regarded as a first step towards building a complete
theory of neutrino flavour in the framework of flavour deconstruction,
and in particular a significant step forward with respect to the hypothesis
of anarchy.

\subsection{Lepton mixing}

The structures of the charged lepton and neutrino Yukawa couplings
are similar up to $\mathcal{O}(1)$ coefficients, therefore we expect
both to contribute to the PMNS mixing angles. Taking into account contributions from both charged leptons
and neutrinos, in all generality the PMNS matrix is given by (see our parametrisations
and conventions for the unitary matrices in Appendix~\ref{sec:conv})
\begin{equation}
U_{\mathrm{PMNS}}=U_{e}U_{\nu}^{\dagger}=U_{12}^{e\dagger}U_{13}^{e\dagger}U_{23}^{e\dagger}U_{23}^{\nu}U_{13}^{\nu}U_{12}^{\nu}\,.\label{eq:PMNS_phases2-1-1}
\end{equation}
In Appendix \ref{app:PMNS} we provide fully general formulas for the PMNS mixing
angles which take into account both charged lepton and neutrino contributions.
In our model, given the structures of Eqs.~\eqref{eq:Yukawa_e} and \eqref{eq:Neutrino_textures}, we expect that
the 13 angles are doubly suppressed as $\sin\theta_{13}^{\nu,e}\sim\epsilon_{12}^{L}\epsilon_{23}^{L}$.
As a consequence, the PMNS angle $\theta_{13}$ is dominated by 12 mixing
in the charged lepton sector, $\theta_{13}\sim\mathrm{sin}\theta_{12}^{e}\sim\epsilon_{12}^{L}\sim0.1$
(see Eq.~\eqref{eq:s13_general}). Interestingly, this suggests that the PMNS angle $\theta_{13}$
originates from 12 charged lepton mixing of similar size to the Cabibbo
angle in the quark sector. All in all, due to the smallness of the
PMNS angle $\theta_{13}$ and the parametric suppression of 13 angles in
our model, we expand the fully general formulas of Eqs.~\eqref{eq:s23_general}, \eqref{eq:s13_general} and \eqref{eq:s12_general} to linear
order in $\theta_{13}^{\nu,e}$ and $\theta_{12}^{e}$ to achieve
simple formulas for the PMNS mixing angles in our model,
\begin{flalign}
 & s_{23} \, e^{-i\delta_{23}}\approx c_{23}^{e}s_{23}^{\nu}e^{-i\delta_{23}^{\nu}}-c_{23}^{\nu}s_{23}^{e}e^{-i\delta_{23}^{e}}\,,\\
 & \theta_{13} \, e^{-i\delta_{13}}\approx\theta_{13}^{\nu}e^{-i\delta_{13}^{\nu}}-\theta_{13}^{e}C_{23}e^{-i\delta_{13}^{e}}-\theta_{12}^{e}S_{23}e^{-i\delta_{12}^{e}}\,,\\
 & s_{12} \, e^{-i\delta_{12}}\approx s_{12}^{\nu}e^{-i\delta_{12}^{\nu}}+\theta_{13}^{e}S_{23}^{*}c_{12}^{\nu}e^{-i\delta_{13}^{e}}-\theta_{12}^{e}C_{23}^{*}c_{12}^{\nu}e^{-i\delta_{12}^{e}}\,,
\end{flalign}
where $c_{ij}\equiv\cos\theta_{ij}$ and $s_{ij}\equiv\sin\theta_{ij}$, and the complex quantities $S_{23}$ and $C_{23}$ are
defined as
\begin{equation}
C_{23}\equiv c_{23}^{e}c_{23}^{\nu}+s_{23}^{e}s_{23}^{\nu}e^{i(\delta_{23}^{e}-\delta_{23}^{\nu})}\,,
\end{equation}
\begin{equation}
S_{23}\equiv c_{23}^{e}s_{23}^{\nu}e^{-i\delta_{23}^{\nu}}-c_{23}^{\nu}s_{23}^{e}e^{-i\delta_{23}^{e}}\,.
\end{equation}
It is illustrative to include explicitly the expressions for these quantities. This leads to
\begin{flalign}
 s_{23} \, e^{-i\delta_{23}}\approx& \, c_{23}^{e}s_{23}^{\nu}e^{-i\delta_{23}^{\nu}}-c_{23}^{\nu}s_{23}^{e}e^{-i\delta_{23}^{e}}\,,\\
 \theta_{13} \, e^{-i\delta_{13}}\approx& \, \theta_{13}^{\nu}e^{-i\delta_{13}^{\nu}}-\theta_{13}^{e}(c_{23}^{e}c_{23}^{\nu}+s_{23}^{e}s_{23}^{\nu}e^{i(\delta_{23}^{e}-\delta_{23}^{\nu})})e^{-i\delta_{13}^{e}} \nonumber \\
 &-\theta_{12}^{e}(c_{23}^{e}s_{23}^{\nu}e^{-i\delta_{23}^{\nu}}-c_{23}^{\nu}s_{23}^{e}e^{-i\delta_{23}^{e}})e^{-i\delta_{12}^{e}}\,,\\
 s_{12} \, e^{-i\delta_{12}}\approx& \, s_{12}^{\nu}e^{-i\delta_{12}^{\nu}}+\theta_{13}^{e}(c_{23}^{e}s_{23}^{\nu}e^{i\delta_{23}^{\nu}}-c_{23}^{\nu}s_{23}^{e}e^{i\delta_{23}^{e}})c_{12}^{\nu}e^{-i\delta_{13}^{e}} \nonumber \\
 &-\theta_{12}^{e}(c_{23}^{e}c_{23}^{\nu}+s_{23}^{e}s_{23}^{\nu}e^{-i(\delta_{23}^{e}-\delta_{23}^{\nu})})c_{12}^{\nu}e^{-i\delta_{12}^{e}}\,.
\end{flalign}
We observe that the large $\theta_{23}$ (atmospheric) angle
is generated from both charged lepton and neutrino contributions,
which may interfere positively or negatively depending on the value
of the phases $\delta_{23}^{\nu}$ and $\delta_{23}^{e}$. The interference
is maximal for $\delta_{23}^{\nu}-\delta_{23}^{e}=(2n-1)\pi$ and
minimal for $\delta_{23}^{\nu}-\delta_{23}^{e}=2n\pi$ (with $n$
being any integer number), which correspond to the case of real mixing
angles. Within the validity of the EFT used to define the $\epsilon_{ij}^{L}$
parameters, we expect some suppression of the 23 angles $s_{23}^{\nu,e}\sim\epsilon_{23}^{L}$,
but this suppression is mild since $\epsilon_{23}^{L}>\epsilon_{23}^{R}\sim0.06$
in order to reproduce lepton data as discussed before. Therefore,
we envisage that large PMNS $\theta_{23}$ is generated either via
a mild tuning (not worst than 10\%), which may be distributed among
the different dimensionless parameters $a_{23,33}^{\nu,e}$, or by
going beyond the EFT to generate $\epsilon_{23}^{L}\sim\mathcal{O}(1)$
as in~\cite{FernandezNavarro:2024hnv}. 

The $\theta_{13}$ (reactor) angle is generated from 12 mixing
in the charged lepton sector, as $\theta_{13}\sim\theta_{12}^{e}\sim\epsilon_{12}^{L}\sim0.1$,
and receives corrections of $\mathcal{O}(\epsilon_{12}^{L}\epsilon_{23}^{L})$
from the 13 angles $\theta^{e,\nu}_{13}$. Interestingly,
$\theta_{12}^{e}\sim0.1$ and the Cabibbo angle $\theta_{c}\sim0.2$
that generates 12 quark mixing are of similar size.

\nopagebreak
{Finally, the $\theta_{12}$ (solar) angle is mostly generated
from 12 neutrino mixing, $s_{12}\approx s_{12}^{\nu}$, with leading
corrections of $\mathcal{O}(\theta_{12}^{e})$. Within the validity
of the EFT used to define the $\epsilon_{ij}^{L}$ parameters, we
expect a mild suppression of the 12 neutrino mixing angle $s_{12}^{\nu}\sim\epsilon_{12}^{L}\sim0.1$.
We consider this acceptable since the solar angle is $s_{12}\sim0.5$.}

\subsection{Quarks}

\begin{table}
\centering %
\begin{tabular}{lcccc}
\toprule 
Field  & $U(1)_{Y_{1}}$  & $U(1)_{R_{2}}\times U(1)_{(B-L)_{2}/2}$  & $U(1)_{R_{3}}\times U(1)_{(B-L)_{3}/2}$ & $SU(3)_{c}\times SU(2)_{L}$\tabularnewline
\midrule 
$q_{1}$  & $\frac{1}{6}$  & $\mathrm{(0,0)}$  & $\mathrm{(0,0)}$ & $(\mathbf{3,2})$\tabularnewline
$q_{2}$  & 0  & $\mathrm{(0,\frac{1}{6})}$  & $\mathrm{(0,0)}$ & $(\mathbf{3,2})$\tabularnewline
$q_{3}$  & 0  & $\mathrm{(0,0)}$  & $\mathrm{(0,\frac{1}{6})}$ & $(\mathbf{3,2})$\tabularnewline
\midrule 
$u_{1}^{c}$  & $-\frac{2}{3}$  & $\mathrm{(0,0)}$  & $\mathrm{(0,0)}$ & $(\mathbf{\overline{3},1})$\tabularnewline
$u_{2}^{c}$  & 0  & $\mathrm{(-\frac{1}{2},-\frac{1}{6})}$  & $\mathrm{(0,0)}$ & $(\mathbf{\overline{3},1})$\tabularnewline
$u_{3}^{c}$  & 0  & $\mathrm{(0,0)}$  & $\mathrm{(-\frac{1}{2},-\frac{1}{6})}$ & $(\mathbf{\overline{3},1})$\tabularnewline
\midrule 
$d_{1}^{c}$  & $\frac{1}{3}$  & $\mathrm{(0,0)}$  & $\mathrm{(0,0)}$ & $(\mathbf{\overline{3},1})$\tabularnewline
$d_{2}^{c}$  & 0  & $\mathrm{(\frac{1}{2},-\frac{1}{6})}$  & $\mathrm{(0,0)}$ & $(\mathbf{\overline{3},1})$\tabularnewline
$d_{3}^{c}$  & 0  & $\mathrm{(0,0)}$  & $\mathrm{(\frac{1}{2},-\frac{1}{6})}$ & $(\mathbf{\overline{3},1})$\tabularnewline
\midrule 
$\phi_{12}^{q}$  & $-\frac{1}{6}$  & $\mathrm{(0,\frac{1}{6})}$  & $\mathrm{(0,0)}$ & $(\mathbf{1,1})$\tabularnewline
$\phi_{23}^{q}$  & 0  & $\mathrm{(0,-\frac{1}{6})}$  & $\mathrm{(0,\frac{1}{6})}$ & $(\mathbf{1,1})$\tabularnewline
\bottomrule
\end{tabular}

\caption{Field content relevant for the quark sector. $\phi_{12,23}^{q}$ are
scalars while the rest are the three usual generations of chiral quarks.
\label{tab:Model_quarks}}
\end{table}

In the quark sector, our model generates natural mass hierarchies
and small quark mixing as in tri-hypercharge. For completeness we
show the explicit results in this section. With the field content shown in Table~\ref{tab:Model_quarks}, we obtain the following
Yukawa couplings for up-quarks and down-quarks\footnote{Notice that in the 32 and 31 entries of the down sector, the operators
$(\phi_{23}^{q})^{2}q_{3}H_{d}d_{2}^{c}$ and $(\phi_{12}^{q})^{2}(\phi_{23}^{q})^{2}q_{3}H_{d}d_{1}^{c}$
also contribute at similar order in the EFT expansion, but have been
neglected for simplicity.}
\begin{equation}
\mathcal{L}_{u}=\left(\begin{array}{ccc}
q_{1} & q_{2} & q_{3}\end{array}\right)\left(\begin{array}{ccc}
a_{11}^{u}\epsilon_{12}^{R}\epsilon_{23}^{R} & a_{12}^{u}\epsilon_{12}^{q}\epsilon_{23}^{R} & a_{13}^{u}\epsilon_{12}^{q}\epsilon_{23}^{q}\\
a_{21}^{u}\epsilon_{12}^{q}\epsilon_{12}^{R}\epsilon_{23}^{R} & a_{22}^{u}\epsilon_{23}^{R} & a_{23}^{u}\epsilon_{23}^{q}\\
a_{31}^{u}\epsilon_{12}^{q}\epsilon_{12}^{R}\epsilon_{23}^{q}\epsilon_{23}^{R} & a_{32}^{u}\epsilon_{23}^{q}\epsilon_{23}^{R} & a_{33}^{u}
\end{array}\right)\left(\begin{array}{c}
u_{1}^{c}\\
u_{2}^{c}\\
u_{3}^{c}
\end{array}\right)H_{u}+\mathrm{h.c.}\,,
\end{equation}
\begin{equation}
\mathcal{L}_{d}=\left(\begin{array}{ccc}
q_{1} & q_{2} & q_{3}\end{array}\right)\left(\begin{array}{ccc}
a_{11}^{d}\epsilon_{12}^{R}\epsilon_{23}^{R} & a_{12}^{d}\epsilon_{12}^{q}\epsilon_{23}^{R} & a_{13}^{d}\epsilon_{12}^{q}\epsilon_{23}^{q}\\
a_{21}^{d}\epsilon_{12}^{q}\epsilon_{12}^{R}\epsilon_{23}^{R} & a_{22}^{d}\epsilon_{23}^{R} & a_{23}^{d}\epsilon_{23}^{q}\\
a_{31}^{d}\epsilon_{12}^{q}\epsilon_{12}^{R}\epsilon_{23}^{q}\epsilon_{23}^{R} & a_{32}^{d}\epsilon_{23}^{q}\epsilon_{23}^{R} & a_{33}^{d}
\end{array}\right)\left(\begin{array}{c}
d_{1}^{c}\\
d_{2}^{c}\\
d_{3}^{c}
\end{array}\right)H_{d}+\mathrm{h.c.}\,,
\end{equation}
where we have defined $\epsilon_{12}^{q}=\langle\phi_{12}^{q}\rangle/\Lambda_{12}$
and $\epsilon_{23}^{q}=\langle\phi_{23}^{q}\rangle/\Lambda_{23}$.
The quark Yukawa texture above delivers the following approximate
scalings for the quark mass eigenvalues, 
\begin{equation}
m_{u}\sim\epsilon_{12}^{R}\epsilon_{23}^{R}\langle H_{u}\rangle\,,\quad m_{c}\sim\epsilon_{23}^{R}\langle H_{u}\rangle\,,\quad m_{t}\sim\langle H_{u}\rangle\,,
\end{equation}
\begin{equation}
m_{d}\sim\epsilon_{12}^{R}\epsilon_{23}^{R}\langle H_{d}\rangle\,,\quad m_{s}\sim\epsilon_{23}^{R}\langle H_{d}\rangle\,,\quad m_{b}\sim\langle H_{d}\rangle\,.
\end{equation}
Just like in the charged lepton sector, the scaling with the small
parameters $\epsilon_{12,23}^{R}$ generates a natural hierarchy among
quark masses. Small quark mixing is naturally generated thanks to
the suppression via the $\epsilon_{12,23}^{q}$ parameters. In particular,
in order to explain CKM mixing we obtain
\begin{equation}
\epsilon_{23}^{q}\sim V_{cb}\sim0.04\,,\qquad\qquad\epsilon_{12}^{q}\sim V_{us}\sim0.2\,,
\end{equation}
while $V_{ub}\sim\epsilon_{23}^{q}\epsilon_{12}^{q}$.

\section{Conclusions \label{sec:conclusions}}

Tri-hypercharge, and more generally the hypothesis of flavour deconstruction,
gives a successful and predictive explanation of fermion mass hierarchies
and small quark mixing. However, in the lepton sector, the mechanisms
presented so far generate an anarchic neutrino flavour structure:
the observed neutrino flavour pattern fully originates as an accidental
configuration of $\mathcal{O}(1)$ dimensionless coefficients. This
allows little predictivity about the origin of neutrino mass eigenvalues
or mixing angles; both normal and inverted mass orderings are possible,
there is no reason to have hierarchical neutrino masses, and there
is no particular understanding of why we have two large and one small
mixing angles in the PMNS matrix. 

While anarchy is a valid approach, it
is widely speculated in the literature that a dynamical mechanism might be responsible for the origin of the lepton
flavour structure. For example, such a dynamical mechanism might satisfy the conditions of sequential right-handed neutrino dominance, leading to a natural neutrino mass hierarchy, with the large lepton mixing angles arising from ratios of Yukawa couplings involving particular right-handed neutrinos. Motivated by such considerations, 
we have shown that the minimal tri-hypercharge theory, when extended to include right-handed neutrinos with $B-L$ gauge charges,
satisfies the sequential dominance conditions, but with a new twist: lepton mixing originates from both the neutrino and charged lepton sectors.

In order to obtain suitable analytic results, it was necessary to go beyond the standard sequential dominance results in the literature (see Appendices~\ref{app:charged_leptons} and \ref{app:PMNS}) in order to allow contributions to lepton mixing angles from both neutrino and charged lepton sectors. The key features of the model-independent results are:
\begin{itemize}
\item We use the same parametrisation and conventions for both charged leptons
and neutrinos, and provide fully general formulas for the lepton mixing angles
in Appendix~\ref{app:PMNS}, including all potential neutrino and charged lepton mixing angles and phases.
\item Within this parametrisation and conventions, we have computed novel
formulas for the charged lepton mixing angles and masses at leading order
in the sequential dominance expansion.
\end{itemize}

Armed with these tools, we have shown that, when the hypothesis of tri-hypercharge
is extended to embrace the right-handed neutrino sector, by decomposing the family hypercharges into the corresponding $B-L$ gauge groups,
then the sequential dominance conditions for 
right-handed neutrinos and charged leptons are satisfied naturally.
The atmospheric neutrino mass $m_{3}$ and the mixing angle $\theta_{23}^{\nu}$ are then
determined by the couplings of a dominant right-handed neutrino.
The solar neutrino mass $m_{2}$ and the mixing angle $\theta_{12}^{\nu}$
are further determined by the couplings of a subdominant right-handed neutrino.
Moreover, these mixing angles and mass eigenvalues are described by
simple analytical formulas at leading order in the sequential dominance
expansion. 

In the present approach, the PMNS mixing angles generally receive contributions
from both charged leptons and neutrinos, in contrast to the usual situation where either neutrinos or charged leptons contribute.
In the considered case, the PMNS angle $\theta_{23}$ originates from large contributions
from charged leptons and neutrinos. Interestingly, the PMNS angle $\theta_{13}$
originates mostly from 12 mixing in the charged lepton sector, similar
in size to the Cabibbo angle. Finally, the PMNS angle $\theta_{12}$ is generated
mostly from 12 mixing in the neutrino sector. The present model also explains
naturally the origin of charged fermion mass hierarchies and small quark
mixing, as in standard tri-hypercharge. The EFT framework that we have employed points towards the scales of symmetry breaking being high. Potential high-scale signatures of this theory remain to be explored.

In principle, the present approach leads to the prospect of predictivity in the neutrino sector, 
due to the lepton mixing angles being simple ratios of Yukawa couplings, 
unlike theories with anarchy where such relations are not present. However, in practice, predictivity is limited by the unknown 
Yukawa couplings, which would need to be constrained by some model, as for example the Littlest Seesaw Models\footnote{See~\cite{Costa:2023bxw} for a recent phenomenological discussion and references to explicit models.}. 
Even without such constrained sequential dominance, we find it remarkable that all fermion mass hierarchies, including that of neutrinos, as well as the disparate lepton and quark mixing patterns, can be qualitatively understood without relying on family symmetry, within a theory of gauge flavour deconstruction.

In summary, we have shown how a natural neutrino mass hierarchy with large lepton mixing angles may be achieved in a theory of gauge flavour deconstruction. The particular framework we considered is based on a tri-hypercharge gauge theory, extended to include 
$B-L$ gauge groups, where we found that the sequential dominance conditions arise naturally, as a consequence of the charged lepton mass hierarchy, and lepton mixing originates from both the neutrino and charged lepton sectors. We have presented new model-independent sequential dominance results applicable to this case but which may also be useful more generally. Finally we showed how natural quark mass and mixing may be included in such a framework.

%Despite this shortcoming, common to all such theories so far, this approach motivates further model building towards the goal of predictivity in the neutrino sector with gauge flavour deconstruction.

%In summary, we show how a natural neutrino mass hierarchy with large lepton mixing angles may be achieved in a theory of gauge flavour deconstruction. In particular we show how minimal tri-hypercharge, suitably extended to include two right-handed neutrinos, together with the charged lepton mass hierarchy, satisfies the conditions of sequential dominance which are a consequence of the charged lepton mass hierarchy
%suitably generalised to include large charged lepton contributions in a model-independent way. 

\section*{Acknowledgements}
MFN and SFK would like to thank the CERN Theory group for hospitality and financial support during an intermediate stage of this work. MFN is supported by the STFC under grant ST/X000605/1. SFK acknowledges the STFC Consolidated Grant ST/X000583/1 and the European Union's Horizon 2020 Research and Innovation programme under Marie Sklodowska-Curie grant agreement HIDDeN European ITN project (H2020-MSCA-ITN-2019//860881-HIDDeN). AV acknowledges financial support from the Spanish grants PID2023-147306NB-I00, CNS2024-154524 and CEX2023-001292-S (MICIU/AEI/10.13039/501100011033), as well as from CIPROM/2021/054 (Generalitat Valenciana).

\appendix

\section{Conventions}
\label{sec:conv}

Note that instead of 4-component left-handed spinors $e_{L},\nu_{L}$
and right-handed spinors $e_{R},\nu_{R}$, we choose to work with
2-component left-handed spinors $e,\nu$ and CP-conjugate right-handed
spinors $e^{c},\nu^{c}$. In this notation, all spinors are left-handed
by construction, and one can drop the chiral indexes. The same notation
applies to the quark sector as well.

In the following, we exhibit the equivalence of different conventions
used to parametrise the PMNS and the respective neutrino and charged
lepton mixing matrices. In all cases we follow Appendix A in \cite{King:2002nf}
for conventions and definitions. In the context of sequential dominance,
the following convention for the PMNS is widely used,\footnote{Note that our convention differs from the usual PDG convention \cite{PDG:2024cfk} where
$V_{\mathrm{PMNS}}=V_{e}^{\dagger}V_{\nu}$.}
\begin{equation}
V_{\mathrm{PMNS}}=V_{e}V_{\nu}^{\dagger}\,.
\end{equation}
With this convention, it is convenient to define the parametrisation
of $V^{\dagger}$ rather than $V$ because the PMNS matrix involves
$V_{\nu}^{\dagger}$ and the neutrino mixing angles typically play
a central role,
\begin{equation}
V^{\dagger}=P_{2}R_{23}R_{13}P_{1}R_{12}P_{3} \, . \label{eq:Parametrisation1}
\end{equation}
Here $R_{ij}$ are a sequence of real rotations corresponding to
real and positive angles $\theta_{ij}$, and $P_{i}$ are diagonal
phase matrices. Note that a $3\times3$ unitary matrix may be parametrised
by 3 angles and 6 phases. Our conventions are as follows
\begin{equation}
R_{23}=\left(\begin{array}{ccc}
1 & 0 & 0\\
0 & c_{23} & s_{23}\\
0 & -s_{23} & c_{23}
\end{array}\right)\,,\quad R_{13}=\left(\begin{array}{ccc}
c_{13} & 0 & s_{13}\\
0 & 1 & 0\\
-s_{13} & 0 & c_{13}
\end{array}\right)\,,\quad R_{12}=\left(\begin{array}{ccc}
c_{12} & s_{12} & 0\\
-s_{12} & c_{12} & 0\\
0 & 0 & 1
\end{array}\right)\,,
\end{equation}
\begin{equation}
P_{1}=\left(\begin{array}{ccc}
1 & 0 & 0\\
0 & e^{i\chi} & 0\\
0 & 0 & 1
\end{array}\right)\,,\quad P_{2}=\left(\begin{array}{ccc}
1 & 0 & 0\\
0 & e^{i\phi_{2}} & 0\\
0 & 0 & e^{i\phi_{3}}
\end{array}\right)\,,\quad P_{3}=\left(\begin{array}{ccc}
e^{i\omega_{1}} & 0 & 0\\
0 & e^{i\omega_{2}} & 0\\
0 & 0 & e^{i\omega_{3}}
\end{array}\right)\,, \label{eq:phases_def}
\end{equation}
where $c_{ij}\equiv\cos\theta_{ij}$ and $s_{ij}\equiv\sin\theta_{ij}$.
By commuting the phase matrices to the left, it can be shown that
the parametrisation in Eq.~\eqref{eq:Parametrisation1} is equivalent
to
\begin{equation}
U^{\dagger}=P \, U_{23}U_{13}U_{12}\,,\label{eq:Parametrisation2}
\end{equation}
where $P=P_{1}P_{2}P_{3}$ and
\begin{equation}
U_{23}=\left(\begin{array}{ccc}
1 & 0 & 0\\
0 & c_{23} & s_{23}e^{-i\delta_{23}}\\
0 & -s_{23}e^{i\delta_{23}} & c_{23}
\end{array}\right)\,,\; U_{13}=\left(\begin{array}{ccc}
c_{13} & 0 & s_{13}e^{-i\delta_{13}}\\
0 & 1 & 0\\
-s_{13}e^{i\delta_{13}} & 0 & c_{13}
\end{array}\right)\,,\; U_{12}=\left(\begin{array}{ccc}
c_{12} & s_{12}e^{-i\delta_{12}} & 0\\
-s_{12}e^{i\delta_{12}} & c_{12} & 0\\
0 & 0 & 1
\end{array}\right)\,,
\end{equation}
where the phases of the $U_{ij}$ matrices are related to the phases
of the $P_{i}$ matrices as follows,
\begin{equation}
\delta_{23}=\chi+\omega_{2}-\omega_{3}\,,\quad\delta_{13}=\omega_{1}-\omega_{3}\,,\quad\delta_{12}=\omega_{1}-\omega_{2}\,.
\end{equation}
We find the parametrisation of Eq.~(\ref{eq:Parametrisation1}) convenient
to diagonalise the neutrino and charged lepton mass matrices, in order
to obtain expressions for the mixing angles and mass eigenvalues including the phases.
In contrast, for the PMNS angles and phases we find more convenient the parametrisation
of Eq.~(\ref{eq:Parametrisation2}), i.e.
\begin{align}
\label{eq:PMNS_convention}
U_{\mathrm{PMNS}}&=P \, U_{23}U_{13}U_{12} \\
&=\left(\begin{array}{ccc}
c_{12}c_{13} & s_{12}c_{13}e^{-i\delta_{12}} & s_{13}e^{-i\delta_{13}}\\
s_{12}c_{23}e^{i\delta_{12}}-c_{12}s_{23}s_{13}e^{-i(\delta_{23}-\delta_{13})} & c_{12}c_{23}-s_{23}s_{23}s_{13}e^{-i(\delta_{12}+\delta_{23}-\delta_{13})} & s_{23}c_{13}e^{-i\delta_{23}}\\
-s_{12}s_{23}e^{i(\delta_{12}+\delta_{23})}-c_{12}c_{23}s_{13}e^{i\delta_{13}} & -c_{12}s_{23}e^{i\delta_{23}}-s_{12}c_{23}s_{13}e^{-i(\delta_{12}-\delta_{13})} & c_{23}c_{13}
\end{array}\right)\,, \nonumber \label{eq:PMNS_convention}
\end{align}
where in the second step the matrix $P$ on the left (and an overall minus sign) have been removed by charged lepton field redefinitions\footnote{Note that this is always possible since right-handed charged lepton
phase rotations can always make the charged lepton masses real.}
$\Delta V^{e}=-P^{\dagger}$. Finally, an alternative is given by 
\begin{equation}
U_{\mathrm{PMNS}}=R_{23}U_{13}R_{12}P_{0}\,,\label{eq:Parametrisation3}
\end{equation}
where $P_{0}=\mathrm{diag}(1,e^{i\beta_{1}},e^{i\beta_{2}})$ and
the phase $\delta_{13}$ in $U_{13}$ is replaced by the Dirac phase
$\delta_{\mathrm{CP}}$, while the phases $\beta_{1}$ and $\beta_{2}$
are known as Majorana phases. This parametrisation is widely used
in the literature in the context of neutrino oscillations, see e.g.~the PDG \cite{PDG:2024cfk}. The parametrisation in Eq.~(\ref{eq:Parametrisation3})
can be transformed to that in Eq.~(\ref{eq:PMNS_convention}) by
commuting the phase matrix $P_{0}$ to the left and then removing
the phases of the left-hand side via charged lepton field redefinitions.
The two parametrisations are then related by the phase relations
\begin{equation}
\delta_{23}=\beta_{2}\,,\quad\delta_{13}=\delta_{\mathrm{CP}}+\beta_{1}\,,\quad\delta_{12}=\beta_{1}-\beta_{2}\,.
\end{equation}

When taking into account contributions from both charged leptons and
neutrinos, the general expression for the PMNS is simplest when
using the parametrisation of Eq.~(\ref{eq:Parametrisation2}) for
the neutrino and charged lepton mixing matrices,
\begin{equation}
U_{\mathrm{PMNS}}=U_{e}U_{\nu}^{\dagger}=U_{12}^{e\dagger}U_{13}^{e\dagger}U_{23}^{e\dagger}U_{23}^{\nu}U_{13}^{\nu}U_{12}^{\nu}\,.\label{eq:PMNS_phases2}
\end{equation}
This is why we prefer to describe the PMNS with the parametrisation
of Eq.~(\ref{eq:Parametrisation2}) rather than with Eq.~(\ref{eq:Parametrisation1})
or Eq.~(\ref{eq:Parametrisation3}). Note that the expression above
has 6 phases, so 3 of them will be removed with charged lepton field
redefinitions to match the expression in Eq.~(\ref{eq:PMNS_convention}). 

The relation between the neutrino and charged
lepton phases that appear in the PMNS and those which are relevant for the
diagonalisation of the mass matrices is as follows,
\begin{equation}
\delta_{12}^{\nu}=\omega_{1}^{\nu}-\omega_{2}^{\nu}\,,\quad\delta_{13}^{\nu}=\omega_{1}^{\nu}-\omega_{3}^{\nu}\,,\quad\delta_{23}^{\nu}=\chi^{\nu}+\omega_{2}^{\nu}-\omega_{3}^{\nu}\,.
\end{equation}
\begin{flalign}
 & \delta_{12}^{e}=\chi^{e}+\phi_{2}^{e}-\phi_{2}^{\text{\ensuremath{\nu}}}-\chi^{\nu}+\omega_{1}^{\nu}-\omega_{2}^{\nu}\,,\\
 & \delta_{13}^{e}=\phi_{3}^{e}-\phi_{3}^{\text{\ensuremath{\nu}}}+\omega_{1}^{\nu}-\omega_{3}^{\nu}\,,\\
 & \delta_{23}^{e}=-\phi_{2}^{e}+\phi_{3}^{e}+\phi_{2}^{\nu}-\phi_{3}^{\nu}+\chi^{\nu}+\omega_{2}^{\nu}-\omega_{3}^{\nu}\,.
\end{flalign}

\section{Sequential dominance formalism}
\label{sec:seq}

Sequential dominance was originally proposed as an elegant and natural
way of accounting for a neutrino mass hierarchy and two large mixing
angles, in contrast to the less appealing idea that these originate
from an anarchic framework. The idea of sequential dominance is that
one of the right-handed neutrinos contributes dominantly to the seesaw
mechanism and determines the atmospheric neutrino mass and mixing.
A second right-handed neutrino contributes subdominantly and determines
the solar neutrino mass and mixing. The third right-handed neutrino
is effectively decoupled from the seesaw mechanism. Subsequently,
this concept has been extended to embrace the charged lepton sector
as well, which may contribute equally to the origin of the PMNS mixing
angles. 

Ultimately, sequential dominance also delivers a simple framework
to study the flavour structure of the lepton sector, where simple
but accurate analytic formulas for mixing angles and mass eigenvalues
are obtained in the form of a perturbative series expansion.

\subsection{Neutrinos \label{app:neutrinos}}

The mechanism of sequential dominance is most simply described by assuming
three right-handed neutrinos in the basis where the right-handed neutrino
mass matrix is diagonal, although it can be also developed in other
bases \cite{King:1998jw,King:1999mb}. In this basis, we parametrise the Dirac mass matrix and the Majorana mass matrix of right-handed
neutrinos as (see e.g.~\cite{King:1998jw,King:1999mb,King:2002nf,Antusch:2004gf})
\begin{equation}
m_{D}=\left(\begin{array}{ccc}
a' & a & d\\
b' & b & e\\
c' & c & f
\end{array}\right)\,,\qquad M_{\mathrm{M}}=\left(\begin{array}{ccc}
X' & 0 & 0\\
0 & X & 0\\
0 & 0 & Y
\end{array}\right)\,,
\end{equation}
where each right-handed neutrino couples to a column in $m_{D}$.
By applying the seesaw formula, we obtain the effective active neutrino
mass matrix as,\footnote{For simplicity we neglect contributions from the decoupled neutrino $X'$ when writing $m_{\nu}$.}
\begin{equation}
m_{\nu}\simeq m_{D} \, M_{\mathrm{M}}^{-1} \, m_{D}^{\mathrm{T}}=\left(\begin{array}{ccc}
\frac{a^{2}}{X}+\frac{d^{2}}{Y} & \frac{ab}{X}+\frac{de}{Y} & \frac{ac}{X}+\frac{df}{Y}\\
. & \frac{b^{2}}{X}+\frac{e^{2}}{Y} & \frac{bc}{X}+\frac{ef}{Y}\\
. & . & \frac{c^{2}}{X}+\frac{f^{2}}{Y}
\end{array}\right) \, ,
\end{equation}
which is symmetric by construction. Sequential dominance occurs when
the right-handed neutrinos dominate the effective neutrino mass matrix
sequentially. This translates to the following dominance prescription
over the model parameters, 
\begin{equation}
\frac{|e|^{2},|f|^{2},|e \, f|}{Y}\gg\frac{|x \, y|}{X}\gg\frac{|x'y'|}{X'}\,,\label{eq:SD_nu}
\end{equation}
where $x,y=a,b,c$, and $x',y'=a',b',c'$. This prescription naturally
delivers normal ordering for the neutrino mass eigenvalues along with a natural neutrino mass hierarchy, 
\begin{equation}
m_{3}^{2}\gg m_{2}^{2}\gg m_{1}^{2}\,.
\end{equation}
Without loss of generality, we have chosen the prescription such that
$Y$ is the dominant right-handed neutrino, with $X$ and $X'$ being
sequentially subdominant. The effective neutrino mass matrix $m_{\nu}$
must be diagonalised by applying a series of unitary transformations,
\begin{equation}
V_{\nu} \, m_{\nu} \, V_{\nu}^{\mathrm{T}}=\mathrm{diag}(m_{1},m_{2},m_{3})\,,
\end{equation}
where, by convention, we choose to parametrise $V_{\nu}$ via three
subsequent 23, 13 and 12 unitary matrices as described in Appendix~\ref{sec:conv}. The prescription of sequential dominance allows to simplify this
diagonalisation process by noting that the contributions associated
to $Y$ are dominant, and those associated to $X$ and $X'$ can be
treated as small perturbations. Therefore, sequential dominance delivers simple approximate formulas for the flavour parameters
in the neutrino sector at leading order in the sequential dominance expansion \cite{King:2002nf}. In the following, we will neglect contributions from the small $\theta^{\nu}_{13}$ angle to show compact formulas.

For the 23 mixing angle we have
\begin{equation}
\tan\theta_{23}^{\nu}\simeq\frac{e}{f} \, e^{-i(\phi_{2}^{\nu}-\phi_{3}^{\nu})}=\frac{|e|}{|f|}\,,
\end{equation}
where the phases $\phi_{2,3}^{\nu}$, which originate from the unitary
rotations (see Appendix~\ref{sec:conv}, Eq.~\eqref{eq:phases_def}), are fixed
as $\phi_{2}^{\nu}-\phi_{3}^{\nu}=\phi_{e}^{\nu}-\phi_{f}^{\nu}$
to make the angle real and positive. 

For the 13 mixing angle we have
\begin{equation}
\theta^{\nu}_{13}\simeq\frac{Y}{|e|^{2}+|f|^{2}}e^{i(\phi_{2}^{\nu}-2\phi_{e}^{\nu})}\left[\frac{a(s_{23}^{\nu}b+c_{23}^{\nu}ce^{i(\phi_{e}^{\nu}-\phi_{f}^{\nu})})}{X}+e^{i\phi_{e}^{\nu}}\frac{d\sqrt{|e|^{2}+|f|^{2}}}{Y}\right]\,.\label{eq:theta13nu_general}
\end{equation}
In our model of Section~\ref{sec:UVmodel}, beyond the sequential dominance conditions
of Eq.~\eqref{eq:SD_nu}, we find the following conditions to be satisfied,
\begin{equation}
\frac{|de|}{Y},\frac{|df|}{Y}\gg\frac{|ab|}{X},\frac{|bc|}{X}\,.
\end{equation}
In this case, the second term in Eq.~\eqref{eq:theta13nu_general} dominates, and we obtain the simpler result
\begin{equation}
\theta_{13}^{\nu}\simeq\frac{d}{\sqrt{|e|^{2}+|f|^{2}}}e^{i(\phi_{2}^{\nu}-\phi_{e}^{\nu})}=\frac{|d|}{\sqrt{|e|^{2}+|f|^{2}}}\,,
\end{equation}
where the phase $\phi_{2}^{\nu}=\phi_{e}^{\nu}-\phi_{d}^{\nu}$ is
fixed to make $\theta_{13}^{\nu}$ real and positive. Together with
the previous condition, this fixes also
$\phi_{3}^{\nu}=\phi^{\nu}_{f}-\phi^{\nu}_{d}$. 

The 12 mixing angle is given by 
\begin{equation}
\tan\theta^{\nu}_{12}\simeq\frac{a}{c_{23}^{\nu}b-s_{23}^{\nu}ce^{i(\phi_{e}^{\nu}-\phi_{f}^{\nu})}}e^{i(\phi_{2}^{\nu}+\chi^{\nu})}=\frac{|a|}{c_{23}^{\nu}|b|\cos\tilde{\phi}_{b}^{\nu}-s_{23}^{\nu}|c|\cos\tilde{\phi}_{c}^{\nu}}\,,\label{eq:theta12nu}
\end{equation}
where the phase $\chi^{\nu}$ originates as well from the unitary
rotations (see Eq.~(\ref{eq:phases_def})), and is fixed
to make $\theta_{12}^{\nu}$ real and positive as
\begin{equation}
c_{23}^{\nu}|b|\sin(\tilde{\phi}_{b}^{\nu})=s_{23}^{\nu}|c|\sin(\tilde{\phi}_{c}^{\nu})\,,
\end{equation}
where we have defined
\begin{equation}
\tilde{\phi}_{b}^{\nu}=\phi_{b}^{\nu}-\phi_{a}^{\nu}-\phi_{2}^{\nu}-\chi^{\nu}\,,
\end{equation}
\begin{equation}
\tilde{\phi}_{c}^{\nu}=\phi_{c}^{\nu}+\phi_{e}^{\nu}-\phi_{f}^{\nu}-\phi_{a}^{\nu}-\phi_{2}^{\nu}-\chi^{\nu}\,.
\end{equation}

Finally, the mass eigenvalues are
\begin{flalign}
m_{3} & \simeq\frac{|e|^{2}+|f|^{2}}{Y}e^{2i(\phi_{e}^{\nu}-\phi_{2}^{\nu}-\omega_{3}^{\nu})}=\frac{|e|^{2}+|f|^{2}}{Y}\,,\label{eq:m3}\\
m_{2} & \simeq\left[\frac{a^{2}}{X}+e^{-2i(\phi_{2}^{\nu}+\chi^{\nu})}\frac{(c_{23}^{\nu}b-s_{23}^{\nu}ce^{i(\phi_{e}^{\nu}-\phi_{f}^{\nu})})^{2}}{X}\right]e^{-2i\omega_{2}^{\nu}}\label{eq:m2}\\
 & =\left[\frac{|a|^{2}}{X}+\frac{(c_{23}^{\nu}|b|\cos\tilde{\phi}_{b}^{\nu}-s_{23}^{\nu}|c|\cos\tilde{\phi}_{c}^{\nu})^{2}}{X}\right]e^{2i(\phi_{a}^{\nu}-\omega_{2}^{\nu})}\nonumber \\
 & =\frac{|a|^{2}}{X(s_{12}^{\nu})^{2}}\,,\nonumber \\
m_{1} & \simeq0
\end{flalign}
where in the second and third steps
of Eq.~(\ref{eq:m2}) we have used Eq.~(\ref{eq:theta12nu}). The phases $\omega_{3}^{\nu}$ and $\omega_{2}^{\nu}$, which
originate as well from the unitary rotations (see Eq.~(\ref{eq:phases_def})),
have been fixed to make the mass eigenvalues real as $\omega_{3}^{\nu}=\phi_{e}^{\nu}-\phi_{2}^{\nu}$
and $\omega_{2}^{\nu}=\phi_{a}^{\nu}$. With this, all the phases from the unitary matrices in the neutrino sector have been fixed to obtain three real and positive mixing angles and mass eigenvalues.

These results show that in sequential dominance the atmospheric neutrino
mass $m_{3}$ and the mixing angle $\theta_{23}^{\nu}$ are determined
by the couplings of the dominant right-handed neutrino with mass $Y$.
The solar neutrino mass $m_{2}$ and the mixing angle $\theta_{12}^{\nu}$
are determined by the couplings of the subdominant right-handed neutrino
of mass $X$. The third right-handed neutrino of mass $X'$ is effectively
decoupled from the seesaw mechanism and leads to the vanishingly small
mass $m_{1}$, in good agreement with the current bounds on the neutrino
scale $\sum m_{\nu}$ by cosmological observations \cite{DESI:2024mwx} and by the KATRIN
experiment \cite{KATRIN:2024cdt}.

We also note that the neutrino mixing angles quoted above correspond
to the unitary matrices that diagonalise $m_{\nu}$. They are equivalent
to the physical angles of the PMNS matrix only when charged lepton
mixing is neglected. Otherwise, in full generality, one obtains the
PMNS matrix as, 
\begin{equation}
V_{\mathrm{PMNS}}=V_{e}V_{\nu}^{\dagger}\,,
\end{equation}
where the charged lepton contribution is described in the next section.

\subsection{Charged leptons} \label{app:charged_leptons}

Sequential dominance in the context of charged lepton mixing generating
the PMNS has been considered in the literature (see e.g.~\cite{Antusch:2004re,King:2005bj,Antusch:2005kw} for dedicated studies), but all
these studies assume that neutrino mixing is subleading, and work with a parametrisation different to the one used when neutrinos dominate. 
Since we are interested
in the case where both charged leptons and neutrinos may contribute
significantly to the PMNS, we need to extend the results to allow for this so that we can treat neutrino and charged lepton contributions consistently.
We therefore obtain new expressions for
the mixing angles with respect to what is found in the literature, in particular using the same parametrisation of unitary matrices that is employed in the neutrino sector.

We define the charged lepton mass matrix as 
\begin{equation}
m_{e}=\left(\begin{array}{ccc}
a' & a & d\\
b' & b & e\\
c' & c & f
\end{array}\right)\,,
\end{equation}
with the SD condition 
\begin{equation}
\left|d\right|,\left|e\right|,\left|f\right|\gg\left|a\right|,\left|b\right|,\left|c\right|\gg\left|a'\right|,\left|b'\right|,\left|c'\right|\,.\label{eq:SD_charged_lepton}
\end{equation}
This predicts hierarchical charged lepton masses and subleading right-handed
charged lepton mixing with respect to left-handed charged lepton mixing.
In the calculation, we include right-handed mixing and the associated
phases, although in the results presented here the right-handed mixing
angles are neglected. Our parametrisation and conventions are shown
in Appendix~\ref{sec:conv}, including the definition of mixing angles and phases.
The general diagonalisation of the charged lepton mass matrix is done by following the recipe of Appendix E in \cite{King:2002nf}. In the following, we show compact results at leading order in the sequential dominance
expansion and neglecting small corrections of order $\theta^{e}_{13}$.

For the 23 mixing angle we obtain,
\begin{equation}
\tan\theta_{23}^{e}\simeq\frac{e}{f} \, e^{-i(\phi_{2}^{e}-\phi_{3}^{e})}=\frac{\left|e\right|}{\left|f\right|}\,,
\end{equation}
where the phases $\phi_{2,3}^{e}$, which originate from the unitary
rotations (see Eq.~(\ref{eq:phases_def})), are fixed
as $\phi_{2}^{e}-\phi_{3}^{e}=\phi_{e}^{e}-\phi_{f}^{e}$ to make
the angle real and positive. 

For the 13 mixing angle we obtain, in
the small angle limit
\begin{flalign}
\theta_{13}^{e} & \approx\frac{d}{c_{23}^{e}fe^{-i\phi_{3}^{e}}+s_{23}^{e_{L}}ee^{-i\phi_{2}^{e}}}=\frac{|d|}{c_{23}^{e}|f|+s_{23}^{e}|e|}\simeq \frac{|d|}{\sqrt{|f|^{2}+|e|^{2}}}\,.
\end{flalign}
where we have fixed the value of the phase $\phi_{3}^{e}$ to make
$\theta_{13}^{e}$ real and positive, $\phi_{3}^{e}=\phi_{f}^{e}-\phi_{d}^{e}$, or equivalently $\phi_{2}^{e}=\phi_{e}^{e}-\phi_{d}^{e}$ from the condition $\phi_{2}^{e}-\phi_{3}^{e}=\phi_{e}^{e}-\phi_{f}^{e}$. 

For the 12 mixing angle we obtain 
\begin{flalign}
\tan\theta_{12}^{e} & \simeq\frac{ae^{i(\phi_{2}^{e}+\chi^{e})}}{c_{23}^{e}b-s_{23}^{e}ce^{i(\phi_{2}^{e}-\phi_{3}^{e})}}=\frac{\left|a\right|}{c_{23}^{e}|b|\cos(\tilde{\phi}_{b}^{e})-s_{23}^{e}|c|\cos(\tilde{\phi}_{c}^{e})}\,,
\end{flalign}
where the phase $\chi^{e}$ originates as well from the unitary rotations
(see Eq.~(\ref{eq:phases_def})), and we have defined
\begin{equation}
\tilde{\phi}_{b}^{e}=\phi_{b}^{e}-\phi_{a}^{e}-\phi_{2}^{e}-\chi^{e}\,,
\end{equation}
\begin{equation}
\tilde{\phi}_{c}^{e}=\phi_{c}^{e}+\phi_{e}^{e}-\phi_{f}^{e}-\phi_{a}^{e}-\phi_{2}^{e}-\chi^{e}\,,
\end{equation}
and fixed the phase $\chi^{e}$ to make $\tan\theta_{12}^{e}$ real,
\begin{equation}
c_{23}^{e}|b|\sin(\tilde{\phi}_{b}^{e})=s_{23}^{e}|c|\sin(\tilde{\phi}_{c}^{e})\,.
\end{equation}

Finally, the mass eigenvalues are
\begin{flalign}
m_{\tau} & \simeq\sqrt{|e|^{2}+|f|^{2}}e^{i(\phi_{3}^{e^{c}}-\phi_{3}^{e}+\phi_{f}^{e}+\omega_{3}^{e^{c}}-\omega_{3}^{e})}=\sqrt{|e|^{2}+|f|^{2}}\,,\\
m_{\mu} & \simeq\left[c_{12}^{e}c_{23}^{e}b-c_{12}^{e}s_{23}^{e}ce^{i(\phi_{2}^{e}-\phi_{3}^{e})}+s_{12}^{e}ae^{i(\phi_{2}^{e}+\chi^{e})}\right]e^{i(\phi_{2}^{e^{c}}-\phi_{2}^{e}+\chi^{e^{c}}-\chi^{e}+\omega_{2}^{e^{c}}-\omega_{2}^{e})}\\
 & =\frac{|a|}{s_{12}^{e}}e^{i(\phi_{2}^{e^{c}}+\phi_{a}^{e}+\chi^{e^{c}}+\omega_{2}^{e^{c}}-\omega_{2}^{e})}=\frac{|a|}{s_{12}^{e}}\,,\nonumber \\
m_{e} & \simeq\left[a'c_{12}^{e}-b's_{12}^{e}c_{23}^{e}e^{-i(\phi_{2}^{e}+\chi^{e})}+c's_{12}^{e}s_{23}^{e}e^{-i(\phi_{3}^{e}+\chi^{e})}\right]e^{i(\omega_{1}^{e^{c}}-\omega_{1}^{e})}\\
 & =|a'|c_{12}^{e}\cos(\tilde{\phi}_{a'}^{e})-|b'|s_{12}^{e}c_{23}^{e}\cos(\tilde{\phi}_{b'}^{e})+|c'|s_{12}^{e}s_{23}^{e}\cos(\tilde{\phi}_{c'}^{e})\,,\nonumber 
\end{flalign}
where we have defined 
\begin{flalign}
 & \tilde{\phi}_{a'}^{e}=\omega_{1}^{e^{c}}-\omega_{1}^{e}+\phi_{a'}^{e}\,,\\
 & \tilde{\phi}_{b'}^{e}=\omega_{1}^{e^{c}}-\omega_{1}^{e}-\phi_{2}^{e}-\chi^{e}+\phi_{b'}^{e}\,,\\
 & \tilde{\phi}_{c'}^{e}=\omega_{1}^{e^{c}}-\omega_{1}^{e}-\phi_{3}^{e}-\chi^{e}+\phi_{c'}^{e}\,,
\end{flalign}
and we have fixed the phases $\omega_{i}^{e^{c}}$ to make the mass
eigenvalues real,
\begin{flalign}
 & \omega_{3}^{e^{c}}=\omega_{3}^{e}-\phi_{3}^{e^{c}}+\phi_{3}^{e}-\phi_{f}^{e}\,,\\
 & \omega_{2}^{e^{c}}=\omega_{2}^{e}-\phi_{2}^{e^{c}}-\phi_{a}^{e}-\chi^{e^{c}}\,,\\
 & |a'|c_{12}^{e}\sin(\tilde{\phi}_{a'}^{e})=|b'|s_{12}^{e}c_{23}^{e}\sin(\tilde{\phi}_{b'}^{e})-|c'|s_{12}^{e}s_{23}^{e}\sin(\tilde{\phi}_{c'}^{e})\,.
\end{flalign}
The three phases $\omega^{e}_{i}$ remain unfixed from the diagonalisation process and can be used to remove three unphysical phases from the PMNS, see Eq.~\eqref{eq:PMNS_convention}.

\section{General formulas for PMNS mixing angles}
\label{app:PMNS}

In this Appendix we provide fully general formulas for the mixing
angles of the PMNS matrix, taking into account both charged lepton and neutrino
contributions. In full generality, the PMNS matrix is given by 
\begin{equation}
U_{\mathrm{PMNS}}=U_{e}U_{\nu}^{\dagger}=U_{12}^{e\dagger}U_{13}^{e\dagger}U_{23}^{e\dagger}U_{23}^{\nu}U_{13}^{\nu}U_{12}^{\nu}\,,\label{eq:PMNS_phases2-1}
\end{equation}
where we consider the conventions and parametrisation for the unitary
matrices discussed in Appendix~\ref{sec:conv}. Note that the expression above has
6 phases, so 3 of them will be removed from charged lepton field redefinitions
to match the expression in \eqref{eq:PMNS_convention}.

Let us focus first on the $U_{23}^{e\dagger}U_{23}^{\nu}$ product,
\begin{equation}
U_{23}^{e\dagger}U_{23}^{\nu}=\left(\begin{array}{ccc}
1 & 0 & 0\\
0 & C_{23}^{*} & S_{23}\\
0 & -S_{23}^{*} & C_{23}
\end{array}\right)\,,
\end{equation}
where we have defined 
\begin{equation}
C_{23}\equiv c_{23}^{e}c_{23}^{\nu}+s_{23}^{e}s_{23}^{\nu}e^{i(\delta_{23}^{e}-\delta_{23}^{\nu})}\,,
\end{equation}
\begin{equation}
S_{23}\equiv c_{23}^{e}s_{23}^{\nu}e^{-i\delta_{23}^{\nu}}-c_{23}^{\nu}s_{23}^{e}e^{-i\delta_{23}^{e}}\,.
\end{equation}
We can extract the PMNS $\mathrm{sin}\theta_{23}$ parameter from the 23 element
of the PMNS matrix, 
\begin{equation}
(U_{\mathrm{PMNS}})_{23}=s_{23}c_{13}e^{-i\delta_{23}}=s_{12}^{e}e^{i\delta_{12}^{e}}(c_{13}^{e}s_{13}^{\nu}e^{-i\delta_{13}^{\nu}}-c_{13}^{\nu}C_{23}s_{13}^{e}e^{-i\delta_{13}^{e}})+c_{12}^{e}c_{13}^{\nu}S_{23}\,.\label{eq:s23_general}
\end{equation}
Now we extract the PMNS $\theta_{13}$ angle from the 13 element of
the PMNS matrix,
\begin{flalign}
(U_{\mathrm{PMNS}})_{13} & =s_{13}e^{-i\delta_{13}}=c_{12}^{e}\left(c_{13}^{e}s_{13}^{\nu}e^{-i\delta_{13}^{\nu}}-c_{13}^{\nu}C_{23}s_{13}^{e}e^{-i\delta_{13}^{e}}\right)-c_{13}^{\nu}s_{12}^{e}S_{23}e^{-i\delta_{12}^{e}}\,.\label{eq:s13_general}
\end{flalign}
Finally, we extract the PMNS $\mathrm{sin}\theta_{12}$ parameter from the 12
element of the PMNS matrix, 
\begin{flalign}
(U_{\mathrm{PMNS}})_{12}= & \;s_{12}c_{13}e^{-i\delta_{12}}\label{eq:s12_general}\\
= & \;c_{12}^{\nu}\left(S_{23}^{*}c_{12}^{e}s_{13}^{e}e^{-i\delta_{13}^{e}}-C_{23}^{*}s_{12}^{e}e^{-i\delta_{12}^{e}}\right)\nonumber\\
 & +s_{12}^{\nu}e^{-i\delta_{12}^{\nu}}\Bigg(c_{12}^{e}c_{13}^{e}c_{13}^{\nu}+s_{13}^{\nu}e^{i\delta_{13}^{\nu}}\left(S_{23}s_{12}^{e}e^{-i\delta_{12}^{e}}+C_{23}c_{12}^{e}s_{13}^{e}e^{-i\delta_{13}^{e}}\right)\Bigg)\,. \nonumber
\end{flalign}

\section{Scalar potential} \label{app:scalarPotential}

The scalar potential of our model contains the following renormalisable
terms,
\begin{flalign}
V_{\phi} & =\sum_{i}m_{i}^{2}\left|\phi_{i}\right|^{2}+\sum_{i,j}\lambda_{ij}\left|\phi_{i}\right|^{2}\left|\phi_{j}\right|^{2}+\left(\lambda_{q_{12}^{3}L_{12}}(\phi_{12}^{q})^{3}\phi_{12}^{L}+\lambda_{q_{23}^{3}L_{23}}(\phi_{23}^{q})^{3}\phi_{23}^{L}+\mathrm{h.c.}\right)\,,\\
V_{H} & =M_{H_{u}}^{2}|H_{u}|^{2}+M_{H_{d}}^{2}|H_{d}|^{2}+\left(M_{H_{ud}}^{2}H_{u}H_{d}+\mathrm{h.c.}\right)+\lambda_{H_{u}}|H_{u}|^{4}+\lambda_{H_{d}}|H_{d}|^{4}\\
 & +\lambda_{H_{u}H_{d}}|H_{u}|^{2}|H_{d}|^{2}+\lambda_{\widetilde{H}_{u}\widetilde{H}_{d}}(H_{u}H_{d})(\widetilde{H}_{d}\widetilde{H}_{u})+\left(\lambda_{H_{u}H_{u}H_{d}H_{d}}H_{u}H_{u}H_{d}H_{d}+\mathrm{h.c.}\right)\,,\nonumber \\
V_{H\phi} & =\sum_{i}\lambda_{i}^{H\phi}|H_{u}|^{2}\left|\phi_{i}\right|^{2}+\sum_{i}\lambda_{i}^{H\phi}|H_{d}|^{2}\left|\phi_{i}\right|^{2}\,,\\
V_{\chi} & =M_{\chi_{3}}^{2}\left|\chi_{3}\right|^{2}+M_{\chi_{2}}^{2}\left|\chi_{2}\right|^{2}+\lambda_{\chi_{3}}\left|\chi_{3}\right|^{4}+\lambda_{\chi_{2}}\left|\chi_{2}\right|^{4}+\lambda_{\chi_{3,2}}\left|\chi_{3}\right|^{2}\left|\chi_{2}\right|^{2}\,,\\
V_{\chi\phi} & =\sum_{i}\lambda_{\chi_{3}\phi}\left|\chi_{3}\right|^{2}\left|\phi_{i}\right|^{2}+\sum_{i}\lambda_{\chi_{2}\phi}\left|\chi_{2}\right|^{2}\left|\phi_{i}\right|^{2}\,,\\
V_{\chi H} & =\lambda_{\chi_{3}H}\left|\chi_{3}\right|^{2}\left|H_{u,d}\right|^{2}+\lambda_{\chi_{2}H}\left|\chi_{2}\right|^{2}\left|H_{u,d}\right|^{2}\,.
\end{flalign}
Our model contains 5 gauge $U(1)$s which are spontaneously broken
down to the diagonal gauge hypercharge. This breaking is performed
by 8 complex scalar degrees of freedom involving the fields $\phi_{12,23}^{R,L,q}$
and $\chi_{2,3}$. Therefore, one has 8 potential Goldstone modes,
out of which 4 are eaten by heavy $Z'$s. Two of the remaining Goldstone
modes acquire a heavy mass at renormalisable level, proportional to
the non-trivial couplings $\lambda_{q_{12}^{3}L_{12}}$ and $\lambda_{q_{23}^{3}L_{23}}$.
Given that we consider our model an EFT, the two final modes may acquire
mass at non-renormalisable level, e.g.~via the couplings
\begin{equation}
V_{d>4}\supset\frac{c_{R12}}{\Lambda}(\phi_{12}^{R})^{2}(\widetilde{\phi}_{12}^{L})^{2}\chi_{2}+\frac{c_{R23}}{\Lambda^{2}}\chi_{3}(\phi_{23}^{R})^{2}\widetilde{\chi}_{2}(\widetilde{\phi}_{23}^{L})^{2}+\mathrm{h.c.}
\end{equation}
From the operators above, the remaining Goldstone modes will get a mass suppressed by the factors $\langle\chi_{2}\rangle/\Lambda$ and $\langle\chi_{3}\rangle \langle\chi_{2}\rangle/\Lambda^{2}$ with respect to the radial modes. Since the radial modes are heavy (around $10^{12}\,\mathrm{GeV}$ or
more) and the VEVs $\langle\chi_{2,3}\rangle \sim 10^{14}\,\mathrm{GeV}$ are large, then we expect the Goldstone modes to acquire a heavy mass, since $\Lambda$ is not expected to exceed the GUT or Planck scales.

In a realistic UV theory, the Goldstone modes likely get a mass at
tree level or loop level. Given that we expect all couplings (including
Yukawa) of the model to be $\mathcal{O}(1)$, and that the radial
modes are expected to be heavy (as mentioned above), we expect any potential Goldstone mode to be very
heavy as well. The electroweak symmetry breaking then proceeds via
the two Higgs doublets $H_{u,d}$, where potential Goldstone modes
may get heavy masses proportional to non-trivial couplings at renormalisable
level, such as $\lambda_{H_{u}H_{u}H_{d}H_{d}}$.

\providecommand{\href}[2]{#2}\begingroup\raggedright\endgroup

\end{document}